\documentclass[a4paper,fleqn,usenatbib]{mnras}

\usepackage[T1]{fontenc}
\usepackage{ae,aecompl}

\usepackage{graphicx}
\usepackage{gensymb}
\usepackage{amsmath}	
\usepackage{amssymb}	
\usepackage{float}
\usepackage{amsfonts}
\usepackage[utf8]{inputenc}
\usepackage{tensor}

\title[GI in magnetised discs]{Gravitoturbulence in magnetised protostellar discs}

\author[]{
A. Riols,$^{1}$ H. Latter $^{1}$
\\
$^{1}$Department of Applied Mathematics and Theoretical Physics, University of Cambridge, Centre for Mathematical Sciences, \\
Wilberforce Road, Cambridge CB3 0WA, United Kingdom. 
}

\date{Accepted XXX. Received YYY; in original form ZZZ}

\pubyear{2016}

\begin{document}
\label{firstpage}
\pagerange{\pageref{firstpage}--\pageref{lastpage}}
\maketitle

\begin{abstract}
Gravitational instability (GI) features
in several aspects of protostellar disk evolution, most notably in
angular momentum transport, fragmentation, 
and the outbursts exemplified by FU Ori and EX Lupi
systems. The outer regions of protostellar discs
may also be coupled to magnetic fields, which could then
modify the development of GI.
To understand the basic elements of their
interaction,
we perform local 2D ideal and resistive MHD simulations with 
an imposed toroidal field. In the regime of
moderate plasma beta, we find
that the system supports a hot gravito-turbulent state, characterised
by considerable 
magnetic energy and stress and 
a surprisingly large Toomre parameter $Q\gtrsim 10$. 
This result has potential implications for disk structure, vertical thickness,
ionisation, etc.
 Our simulations also reveal the
existence of long-lived and dense `magnetic islands' or plasmoids.
Lastly, we find that the presence of a magnetic field has 
little impact on the fragmentation criterion of the disk. 
Though our focus is on protostellar disks, some of our results may be
relevant for the outer radii of AGN.

\end{abstract}




\section{Introduction}

In the early stages of star formation, protoplanetary discs may be
subject to gravitational instability (GI) owing to their
large densities and low temperatures. 
The parameter that best quantifies a disk's 
susceptibility to GI is the Toomre $Q$,
defined via
\begin{equation}
Q=\dfrac{c_s\kappa}{\pi G \Sigma_0}<1
\label{mass_eq}
\end{equation}
\citep{toomre64}, 
where $c_s$ is
the sound speed, $\kappa$ the epicyclic frequency,
and $\Sigma_0$ the background surface density. 
In a razor thin disk, the linear instability criterion for axisymmetric
disturbances is simply $Q<1$, though
non-axisymmetric nonlinear instability occurs for slightly
larger $Q$. When radiative cooling is inefficient, the system saturates
in a gravitoturbulent state that can transport significant angular
momentum \citep{Gammie2001,rice14}, while more efficient cooling causes
the system to fragment into dense clumps that may serve as the
precursors of gas giant planets \citep{cameron78,boss97}.
Note that the critical cooling time that separates the two outcomes is
vulnerable to the numerical details of its calculation and still the
subject of some debate \citep{Pdkooper2012,rice14}. 
 \\

A key but undeveloped area of research is the interaction between the
GI and magnetic fields, and in particular the magnetorotational instability (MRI),
an alternative mechanism of angular momentum transport
in sufficiently ionized gas \citep{balbus91,hawley95}.
In the massive early stage of a protostellar disk's life, GI
may lead to gravitoturbulence and fragmentation at large radii, 
but these regions could also be
ionised to a dynamically relevant degree by cosmic rays or stellar X-rays 
\citep[e.g.][]{armitage11}. Obvious questions are whether the GI and
MRI coexist and a quasi-steady state accommodating both is possible,
what the properties of this state might be, and if fragmentation is enhanced
(or mitigated) by the MRI. 
Even if the MRI is quenched or
greatly impeded by non-ideal MHD effects, the gas may still couple to 
large-scale magnetic fields \citep[possibly generated by Hall
currents][]{simon15},
 which could then significantly modify the gravitoturbulence.
On the other hand, GI might act as a dynamo, creating
small-scale field from a low-amplitude seed.  \\

The interaction between GI and magnetic fields may also be important
later in a disk's lifetime, during 
FU Ori and EX-Lupi outbursts, when the accretion rate undergoes
violent jumps on timescales of 100-1000 years
\citep{evans09,aguilar12}. 
It is theorised that this quasi-periodic behaviour conforms to a
`gravo-magneto' limit cycle, according to which (a)
mass accumulates in the dead
zone, via more efficient accretion at larger radii, 
until (b) the high surface density initiates GI, which heats the
gas to the point that (c) collisional ionisation permits the onset of 
MRI,
and (d) the excess mass is swept onto the protostar in a dramatic
accretion event \citep{armitage01,zhu10,martin11}. 
One issue here is the strong heating required: can
GI adequately thermalise its turbulent motions so as to trigger the MRI?
Another issue is whether the MRI can emerge unproblematically from the
pre-existing gravitoturbulent state.\\

Global and local simulations of self-gravitating discs have been
extensively used over the last decade but very few have coupled GI
with MHD. \citet{kim2001}
studied the fragmentation criterion in magnetized galactic discs, but
their local simulations do not reproduce a fully saturated
 gravitoturbulent state. 
The coexistence between GI and MRI was
investigated by Fromang et al.~(2004a, 2004b) and Fromang (2005) 
 who showed that the turbulence
induced by MRI modes tends to reduce the strength of the gravitational
instability and prevent local clumps of gas from collapsing. That
being said,
these global simulations (however pioneering) suffered from a lack of
resolution and probably did not adequately capture the characteristic
lengthscales of either instability. On the other hand, numerical
studies of outbursts involving both MRI and GI model one or both
process as a diffusion with an effective alpha parameter (e.g., Armitage et
al.~2001, Zhu et al.~2010). Though limit
cycles can be obtained this way, there is yet no direct evidence that this
is the case when the
different turbulent flows are simulated directly.\\

 One obvious response to these issues is to perform
 3D vertically stratified shearing box simulations in which the
 intrinsic scales of both instabilities are resolved. This is a
 computationally demanding task, as the MRI inhabits lengthscales
 less than the scale height $H$, while the GI saturates on scales
 much greater than $H$. A preliminary (and almost unavoidable)
 approach is to conduct 2D MHD simulations. Though this simpler
 setup precludes the MRI, it allows us to identify
 important MHD processes that should be shared by 3D simulations.
 It also provides a fair description of places in the disk that are
 magnetically active and yet MRI-stable, such as at certain outer
 radii in massive young disks and the dead zones of older disks
 at the onset of
 an outburst. In this paper we present a suite of such 2D
 simulations combining MHD and GI. Our computational domains are, for
 the most part, threaded by a mean toroidal field of varying strengths
 and the gas is endowed with a simple linear cooling law. Both ideal
 and resistive MHD are tested. Though ambipolar diffusion and the Hall effect
 are important (often dominant)
  players in the weakly ionised plasma, they are
 omitted here for simplicity. In our simulations, Ohmic (or grid) diffusion may be
 interpreted as a very crude proxy for whatever
 process is diffusing and destroying magnetic field.      \\

The main result of our exploratory work is that the presence of an imposed magnetic field
can dramatically change the thermodynamic properties of the
gravito-turbulent state. The turbulent motions stretch, distort, and amplify the
magnetic field to strengths of order, or even exceeding, the kinetic
energy. Dissipation of this energy leads the system to
a quasi-steady state that is
markedly hotter than in hydrodynamical simulations, with a mean Toomre
$Q$ sometimes over 10. The mean $Q$ correlates with the
strength of the imposed magnetic field. Adding resistivity weakens
this phenomena but does not qualitatively change the picture. 
The details of the dissipation are striking, with energy
thermalised primarily in current sheets and in the slow shocks
generated by reconnection events. Reconnection also gives rise to
`magnetic islands', or plasmoids, that persist for hundreds of orbits.
Finally, we investigate the propensity of the system to fragment as we
change the cooling time. In summary, no great qualitative change in
the critical cooling time is observed. 

The structure of the paper is as follows.
In the following section we present the model, its governing
equations, and the numerical methods that we deploy in their solution.
Our main ideal MHD results appear in section 3, the principal 
control parameter
being the strength of the imposed field. Effects induced by resistivity are
investigated in section 4. In section 5, we go into more detail
exploring the nature of reconnection in the simulations. 
Finally, in section 6, we discuss the
astrophysical implications of this work and how it prepares for future
simulations in 3D.

\section{Model and numerical framework\label{model}}
\subsection{{Model and equations}}

The physical set-up, governing equations, and numerical approach is similar to that described
by \citet{Pdkooper2012}. We use a local Cartesian model
of an accretion disk \citep[the shearing sheet;][]{goldreich65},
whereby the axisymmetric differential rotation is approximated locally
by a linear shear flow  $\mathbf{u}_0=-S x\,  \mathbf{e}_y$ and a
uniform rotation rate $\boldsymbol{\Omega}=\Omega \, \mathbf{e}_z$, with
$S=(3/2)\,\Omega$ for a Keplerian equilibrium. We denote $(x,y,z)$
respectively as the shearwise, streamwise and spanwise directions,
corresponding to the radial, azimuthal and vertical directions. 
We also refer to the $y$ projection of a vector field as its
toroidal component. We neglect the vertical structure of the disc
and consider it infinitely thin, so that the gas is allowed to
move only in a two-dimensional frame $(z=0)$. For simplicity, we
assume that the gas is ideal, its pressure $P$ and surface
density $\Sigma$ related by $\gamma P=\Sigma c_s^2$, where $c_s$
is the sound speed and $\gamma$ the ratio of specific heats. The
pressure is related to internal energy $U$ by $P=(\gamma-1)U$. The
evolution of surface density $\mathbf{\Sigma}$, velocity {field}
perturbations $\mathbf{u}$, magnetic field $\mathbf{B}$ and internal
energy $U$ is then governed by the 2D compressible  dissipative MHD
equations:
\begin{equation}
\dfrac{\partial \Sigma}{\partial t}+\nabla\cdot \left(\Sigma \mathbf{u}\right)=0,
\label{mass_eq}
\end{equation}
\begin{multline}
\frac{\partial{\mathbf{u}}}{\partial{t}}-Sx\frac{\partial{\mathbf{u}}}{\partial{y}}+\mathbf{u}\cdot\mathbf{\nabla
  u}-Su_x\mathbf{e}_y +2\boldsymbol{\Omega}\times\mathbf{u} =-\nabla\Phi\\
  +\frac{1}{\Sigma}(-\mathbf{\nabla}\mathcal{P}+\mathbf{B}\cdot\mathbf{\nabla B}+\nabla\cdot \boldsymbol{\Pi}),
\label{ns_eq}
\end{multline}
\begin{equation}
\frac{\partial{\mathbf{B}}}{\partial{t}}-Sx\frac{\partial{\mathbf{B}}}{\partial{y}}  = -SB_x\mathbf{e}_y+\nabla\times(\mathbf{u}\times\mathbf{B})+\eta\mathbf{\Delta B},
\label{magnetic_eq} 
\end{equation}
\begin{equation}
\dfrac{\partial U}{\partial t}-Sx\dfrac{\partial U}{\partial
  y}+\nabla\cdot (U\mathbf{u})
  =-P\nabla\cdot\mathbf{u}-\dfrac{U}{\tau_c}+Q_D+\kappa_{th}{\Delta T} .
\label{energy_eq}
\end{equation}
To this set we must add the solenoidal condition $\nabla\cdot\mathbf{B}=0$.
In the Navier Stokes equation \eqref{ns_eq}, $\mathcal{P}$ is the sum of gas
pressure $P$ plus magnetic pressure $B^2/2$ and $\Phi$ is the
gravitational potential induced by the disc, obeying the Poisson
equation. The (molecular) viscous stress tensor is $\sf{\Pi}$ and is
defined by
\begin{equation}
\boldsymbol{\Pi} = \Sigma\nu\left[\nabla\mathbf{u} +
\left(\nabla\mathbf{u}\right)^T-\tfrac{2}{3}(\nabla\cdot\mathbf{u})\boldsymbol{I} \right].
\end{equation}
The constant kinematic viscosity and magnetic diffusivity are denoted
by $\nu$ and $\eta$. In the energy equation, we use a cooling law
that is linear in $U$ and whose typical timescale is  $\tau_c$ (also
called the cooling time). The viscous and Ohmic heating is 
$Q_D=\boldsymbol{\Pi} :  \mathbf{\nabla v} + \eta \vert
\nabla \times \mathbf{B} \vert ^2 $. The last term on the right hand
side of the energy equation describes
thermal
conduction, which involves the temperature $T=P/ (R\Sigma)$, with $R$ the gas
constant, and the thermal conductivity 
$\kappa_{th}$. We define $\Omega^{-1}$ as our
unit of time and $H_0=c_{s_0}/\Omega$ our unit of length where
$c_{s_0}$ 
is the uniform sound speed of the background laminar state at $t=0$.

Lastly, $\Phi$ is computed from the Poisson equation
\begin{equation}
\nabla^2\Phi = 4\pi G\rho,
\label{poisson_eq}
\end{equation}
where $\rho$ is the three-dimensional density distribution of the gas
which may be related to the surface density via 
\begin{equation}
\rho(x,y,z) = \Sigma(x,y)\delta(z),
\label{poisson_eq}
\end{equation}
with $\delta$ the Dirac delta function. Note that we omit
a smoothing length, thus the self-gravitational potential
can have scales comparable to the grid size of the simulations. The
effect of a smoothing length is discussed in \cite{Pdkooper2012}.

\subsection{Diagnostics}
\label{alpha}

First let us define 
$\left<. \right>=\frac{1}{L_xL_y}\int\int
\left(\int_{-\infty}^{\infty}\, .\,\, dz\right) dx dy$ as the volume
average of a quantity over a Cartesian portion of size
$L_x$ and $L_y$. A quantity that will be widely used in this paper is
the coefficient $\alpha$ which measures the
angular momentum transport. This quantity is related to  
the average Reynolds stress $H_{xy}$, Maxwell stress $M_{xy}$, gravitational stress $G_{xy}$ and molecular viscous stress $\Pi_{xy}$ by:
\begin{align}
\alpha=\dfrac{2}{3\gamma \left\langle P\right\rangle}\left\langle
  H_{xy}+M_{xy}+G_{xy}+\Pi_{xy} \right\rangle,
\end{align} 
where
\begin{align*}
H_{xy}=\Sigma u_xu_y \quad M_{xy}=-B_xB_y \quad \text{and} 
\quad G_{xy}=\dfrac{1}{4\pi G}\dfrac{\partial \Phi}{\partial
  x}\dfrac{\partial \Phi}{\partial y}.
\end{align*}
It is straightforward to show that the radial flux of angular momentum
gives rise to the only source of energy in the system that can balance the
cooling. This energy, initially in the form  of kinetic energy, can 
be stored in magnetic fields but is irremediably converted into heat by turbulent motions. 

In order to study the energy budget of the flow, we introduce 
the average kinetic, magnetic,
gravitational and internal energy denoted by
\begin{align*}
 E_c=\frac{1}{2}\langle\Sigma \mathbf{u}^2\rangle, \quad
E_m=\frac{1}{2}\langle\mathbf{B}^2\rangle, \quad E_G=~\langle \Sigma \Phi
+\frac{1}{8\pi G}\vert\mathbf{\nabla \Phi}\vert^2\rangle,
\end{align*}
and $U=(\gamma-1)\langle P \rangle$ respectively. Although the temperature and
thermodynamic balance  can be very different from one cell to
another,
 we define an average Toomre parameter in the domain
\begin{equation}
Q=\dfrac{\left\langle c_s\right\rangle \Omega}{\pi G \left\langle \Sigma\right\rangle }
\end{equation}

\subsection{Numerical methods}

We employ the 2D shearing box to simulate locally the motion of
the fluid. Because
the fluid in a gravito-turbulent disc is compressible and is mostly
heated by shocks, we use the PLUTO code \citep{mignone2007} to
perform direct numerical simulations  of
Eqs. \eqref{mass_eq}-\eqref{poisson_eq}. This code uses a Godunov
scheme, a conservative finite-volume method that solves the
approximate Riemann problem at each inter-cell boundary. This scheme
is known to successfully reproduce the behaviour of conserved quantities
like mass, momentum and energy through discontinuities. The Riemann
problem is handled by the HLL solver which has the advantage of being
robust and preserving positivity. Usually HLLD solvers are more
suitable for MHD problems but we checked that our results are not
strongly modified when the HLL solver is used. In the shearing box
framework, simulations are performed in a finite domain of size
$(L_x,L_y)$, discretised on a mesh of $(N_X,N_Y)$ grid points. The
boundary conditions are periodic in $y$,  
while  shear-periodicity is imposed in $x$.

To compute the gravitational potential, we take advantage of the
shear-periodic boundary conditions, following \citet{Gammie2001}. At
each time step, we first shift back the density in $y$ to the time it
was last periodic $(t = t_p)$. For this, we perform a 1D forward
Fourier transform in $y$ for each $x$, multiply by a complex phase
$\exp{(-\text{i}Sk_yx(t-t_p))}$, and take the inverse 1D Fourier
transform. As the resulting surface density and gravitational
potential are periodic in $x$ and $y$, they can be expressed as a
discrete sum of Fourier modes ($\Sigma_k$, $\Phi_k$) with wavevectors
$\mathbf{k}=(k_x,k_y)$. The Fourier decomposition is done with a 2D
FFT algorithm. We then solve the Poisson equation in Fourier 
space where a solution for a single mode is
\begin{equation}
\Phi_k = -\dfrac{2\pi G\Sigma_k}{|\mathbf{k}|}
\label{fourier_decompo}
\end{equation}
By multiplying $\Phi_k$ by $\text{i}k_x$ and $\text{i}k_y$, we obtain the
self-gravity force in the Fourier space. An inverse
FFT delivers the force in the real domain. 
The linear stability of an infinitely thin layer
has been tested to ensure that our implementation is correct
(see Appendix A). Note that gravitational 
 energy and stresses are computed directly in
Fourier space in the same way as \citet{Gammie2001}.

Finally, we use the orbital advection algorithm of PLUTO, based on
splitting the equation of motion into two parts, the first containing
the linear advection operator due to the background Keplerian shear
and the second the standard MHD fluxes and source terms. This
operation allows larger time steps and eliminates numerical artifacts
at the boundaries where the Mach number associated with 
the background shear flow can be very large.

\subsection{Simulation setup}

\subsubsection{Box size and resolution}

The axisymmetric linear theory for thin discs shows that the flow is
unstable for $Q\leq 1$, with the fastest growing mode possessing
a radial lengthscale of order $2\pi H\,Q$.
Although our simulations are focused on the
regime $Q\gtrsim 1$, we expect typical lengthscales
to be also $\gtrsim H$. In order to
obtain a good statistical average of the fluctuating properties, it is
then necessary that $L_x \sim L_y \gg H$. Our fixed reference
lengthscale is the initial scale height of the gas $H_0$. 
As the gas heats up (or cools down) the temperature, and
hence the the scale height $H$, varies.
 Previous hydrodynamic simulations show that steady turbulent flows are able to
sustain an average $Q$ around $2-3$, which translates to an average
$H\sim 1-2\, H_0$. We hence choose $L_x=L_y=40\,H_0$ to
make sure that the structures that develop in the box are much smaller
than the box size. For comparison \citet{Gammie2001} and
\citet{Pdkooper2012} 
used a box of size 100 $H_0$. 

An appropriate resolution is not easy to guess. 
The work of \citet{Gammie2001} suggests that a resolution of 5 grid
cells per $H_0$ is the minimum required. This
ensures that  the energy lost by the numerical scheme remain small
compared to the energy radiated away by the cooling law. However,
\citet{Pdkooper2012} showed that the
fragmentation criterion is still dependant on resolution when the
latter exceeds 40 points per $H_0$. In particular, increasing
resolution leads to easier fragmentation at higher values of $\tau_c$.
In fact, fragmentation appears to be
a stochastic process whose probability of occurrence decreases with
increasing $\tau_c$. The reasons for this resolution dependence remain
unclear and might depend on the algorithm or code
implementation. \citet{Pdkooper2012} argued that the numerical scheme
and resolution needs to be sufficiently accurate so as to maintain a coherent
clump of size $H$ over many dynamical timescales. In this paper, we used
a resolution of 51 points per height scale $H_0$ which translates to
$N_X=N_Y=2048$ for the entire box so that we are slightly better
resolved than the most accurate run of \citet{Pdkooper2012}.

Because of the prevalence of shocks in the compressible
gravitoturbulence, it is practically impossible to viscously 
resolve the shortest scales. However, we checked that
average turbulent quantities (such as mean $Q$, the mean energies, etc)
remain relatively unchanged when using a resolution  
of $N_X=N_Y=1024$, suggesting that our simulations are resolved in
this respect.
Small-scale magnetic features, on the other hand, are possible to
resolve physically if Rm is sufficiently low.

\subsubsection{Initial conditions}
\label{ic}
Initial conditions require particular attention as they determine
if the flow reaches a steady turbulent state or not. We
start our simulations with a uniform density distribution
$\Sigma_0=1$. The total mass in the box is conserved so that $\left<\Sigma\right>$ at any time is equal to $\Sigma_0$.
The initial Toomre parameter $Q$ cannot be smaller than 1 since linear
axisymetric fluctuations automatically lead to
fragmentation. To make sure that such fluctuations cannot grow, we  
choose $Q_0=1.6$ at $t=0$ which corresponds to a fixed gravitational constant $G=0.2$ in all simulations. 
We generate a random seed in the initial density and velocity
perturbations by injecting a small amount of energy in all $k_x$ and
$k_y$ Fourier components. We find that the noise amplitude has to be
sufficiently high to excite a turbulent flow, which confirms that
the transition to such flow is subcritical. Starting with a 
sufficiently large fluctuation, the development of the turbulence is not
immediate but takes a finite time $t_\text{trans}$. For cooling times
smaller than $t_\text{trans}$, the fluid can cool down to $Q < 1$
before any heating through turbulent motions. This leads to premature
fragmentation. To avoid that, we switch on the
cooling term after  a turbulent state has been reached. 

For MHD simulations, initial velocity and density fields are taken
from a pre-existing gravito-turbulent state obtained by a
hydrodynamic run. A large scale uniform toroidal magnetic field
$B_{y_0}$  is then introduced into the box at $t=0$. We define the
initial beta via
\begin{equation}
\beta_0=\frac{2 \Sigma_0 c_{s_0}^2}{B_{y_0}^2},
\end{equation}
the ratio of gas to magnetic pressure of the background laminar
state. The main results of this study are restricted to the case
$\beta_0>1$. Zeldovich's theorem suggests that
no dynamo action is possible in a 2D model, and so a zero-net flux
field will decay over time.
 However, the
average toroidal magnetic flux
is conserved during our simulations, which allows rms-magnetic
fluctuations to be maintained indefinitely. 

\subsubsection{Cooling and diffusion parameters}
\label{diffusion_coeff}

In our model the total energy is
removed via the term $-U/\tau_c$ which mimics 
radiative cooling with an adjustable timescale $\tau_c$. The validity
of this simple cooling law, and realistic values of $\tau_c$,
are important issues. The cooling time in a protostellar disk varies by
orders of magnitude between different radial locations and at different
stages of a disk's evolution. In the later T-Tauri or class-II stages,
$\tau_c$ can be $<1/\Omega$, but in younger class-0 disks the cooling time
can be considerably longer. \citet{kratter16} compute
representative values of $\tau_c$ for various sources, and deduce that (generally)
$\Omega\tau_c>10$ in massive non-fragmenting disks (their exemplar is 
IRAS 16293-2422b). Following on from this work, we adopt a range
$\Omega\tau_c=1-50$. 

Although
PLUTO conserves total energy, the shearing boundaries do work on the
fluid and provide a source of energy. Depending on how
these source terms are computed, numerical errors can be much larger
than roundoff errors and produce a numerical loss of energy. We checked,
however, that the numerical loss intrinsic to the code is small
compared to physical dissipation. By analysing each
term individually in the global energy budget, we were able to
quantify the ratio between the numerical loss and the total energy
content. We found that for a moderate cooling time this ratio remains
smaller than $10^{-4}$ per dynamical time, which means that on average
less than 10\% of the energy is lost after 1000 $\Omega^{-1}$. In
comparison, \citet{Gammie2001} has a relative numerical  loss of the 
order $10^{-3}$ per dynamical time, estimated from the difference between their numerical and predicted $\alpha$.

Internal exchange of energies are possible, at least in part, through
the action of dissipative processes such as viscous friction or Ohmic
diffusion, which convert kinetic and magnetic energy into
heat irreversibly. In addition, internal energy $U$ is redistributed
through the fluid by thermal diffusivity.  In our simulations, we
introduced a uniform tiny viscosity, such that the Reynolds number
$\text{Re}=\Omega H^2/\nu=1000$, and a moderate thermal conductivity
$\kappa_{th}=0.06$. These coefficients are probably not representative
of any astrophysical disc but avoid large velocity or
temperature gradients. A test has been done with $\kappa_{th}=0$ for
which the average turbulent quantities 
remain quite similar to those with $\kappa_{th}=0.06$. 

Ohmic resistivity is known to play a significant role in the turbulent
dynamics of accretion discs. Its influence on self-gravitating MHD
turbulence will be first neglected in the simulations of section 
\ref{mhd_runs} (so that magnetic energy is dissipated on the grid)
but taken into account in section \ref{resistive_runs}. The magnetic
Reynolds number $\text{Rm}=\Omega H^2/\eta$, defined  as the typical
ratio between the advective term and the resistive term, will be
varied from 10 to 5000. For comparison, a very crude estimate of the numerical grid's magnetic
Reynolds number is $= (H_0/L)^2N_X^2 \approx 2500$, though grid diffusion will
not operate like a Laplacian nor be isotropic.
 Lastly, 
we recognise that ambipolar diffusion and the Hall effect play a
significant and usually dominant role in the external regions of
protoplanetary discs. 
However, given that our work is exploratory, we omit
more complicated non-ideal MHD for simplicity. We hence regard
Ohmic diffusion in our model as a (very) coarse proxy for whatever diffusive
process is dominating locally (which could also include small-scale
magnetic turbulence driven by the MRI). This is discussed in more detail in
section 4.

\section{Gravito-turbulence with and without a magnetic field}

In this section, we present several pure hydrodynamical
simulations that test our code and provide a point of
comparison with later magnetized simulations. We then study the coupling
between the gravito-turbulence and a magnetic field, focussing especially
on global properties  and the fragmentation
criterion in different magnetic regimes, from weakly magnetized 
($\beta_0 \gg 1$) to rather strongly magnetized ($\beta_0 \sim 1$). 

\subsection{Hydrodynamical simulations}
 
In hydrodynamic shearing box simulations with $\tau_c\gg
\Omega^{-1}$, the system settles on
a strongly turbulent state that maintains $Q$ around unity. The
heat generated by turbulent motions acts as a feedback loop that
regulates the thermodynamic state. For example, if $Q$ takes values
too
low, the instability becomes more active and enhances the gas
temperature so that the system returns to equilibrium. This idea was
first
proposed by \citet{paczynsky78} and
numerically demonstrated by \citet{Gammie2001} in the shearing box. The
latter also showed that gravito-turbulence transports a significant
amount of angular momentum and predicted that this transport is
inversely proportional to the cooling time $\tau_c$.  When $\tau_c
\sim \Omega^{-1}$, the behaviour is radically different and the disc
fragments into massive clumps. For $\gamma=2$ and a numerical
resolution of $10$ points per scale height, \citet{Gammie2001} found
that the critical $\tau_{c}$ for which fragmentation occurs is
$\simeq  3 \Omega^{-1}$. In actual fact, there is no
clear transition between sustained gravitoturbulence and fragmentation,
as explained by \citet{Pdkooper2012}. 
Note also that localized fragments can form stochastically
 without disrupting the whole disc.  \\
\begin{figure}
\centering
\includegraphics[width=\columnwidth]{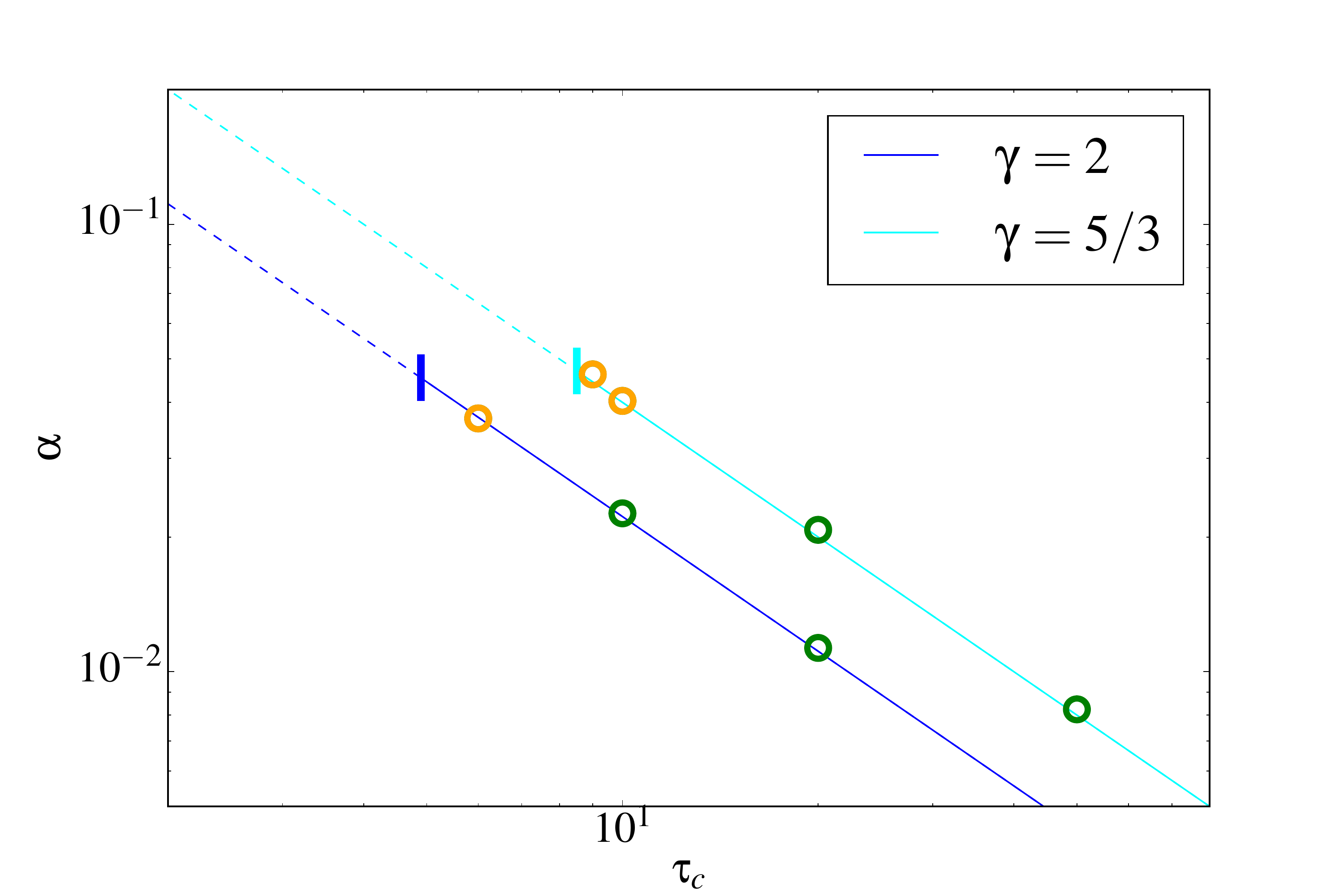}
 \caption{The angular momentum transport coefficient $\alpha$ as a
   function of $\tau_c$ in purely hydrodynamic gravitoturbulent
   simulations. Green circles represent simulations with no fragments
   while orange circles represent simulations where one or several transient fragments are observed. The straight lines are the theoretical predictions of \citet{Gammie2001} and the end bars correspond to the critical cooling time below which the disc is fragmenting as a whole.}
\label{fig_frag_hydro}
 \end{figure}
 
In order to check our code and compare our results with these previous
studies, we performed several simulations without magnetic field
($B_{y_0}=0$) and a varying $\tau_c$. We used the procedure
described in section \ref{ic} to generate suitable initial
conditions. Two different adiabatic indices were considered,
$\gamma=5/3$ and $\gamma=2$. Simulations that did not fragment into one
or few massive clumps were run for at least
$600\,\Omega^{-1}$ in order to obtain well-defined saturated
states. Fig.~\ref{fig_frag_hydro}  shows the average angular momentum 
transport coefficient $\alpha$ of these turbulent states, 
defined in section \ref{alpha}, as a function of $\tau_c$.
We show that for both $\gamma=5/3$ and $\gamma=2$, the $\alpha$
coefficient follows the theoretical prediction of \citet{Gammie2001}
(indicated by straight lines). As this prediction is derived from
an energy conservation principle, this result is just saying that
numerical energy
losses are small in our simulations. Fig.~\ref{fig_frag_hydro} also shows
that these turbulent states collapse into clumps when $\tau_c$
is decreased, though the transition is not
necessarily well defined, as in \citet{Pdkooper2012}. For $\gamma=2$, local and transient
fragments appear first for $\tau_c=6\,\Omega^{-1}$ while the entire
 computational domain fragments when
$\tau_c\lesssim 5\,\Omega^{-1}$. For $\gamma=5/3$, the first fragments 
appear at $\tau_c\lesssim 10\,\Omega^{-1}$ while massive
unstable clumps are formed below $\tau_c\lesssim 8-9 \,\Omega^{-1}$.  
In both cases, the critical $\alpha_c$ for which the disc fragments is comparable and around $\alpha_c \simeq 0.04-0.05$. 

We found that the average Toomre parameter $Q$ in steady turbulent
simulations does not depend strongly on $\gamma$. However it seems to
slightly decrease as $\tau_c$ is reduced, going from $Q=3$ when
$\tau_c=50\,\Omega^{-1}$ to $Q=2$ when $\tau_c=10\,\Omega^{-1}$. This
behaviour is not very surprising: when $\tau_c$ is decreased,
cooling is enhanced requiring a commensurate increase in turbulent
heating by GI, only possible by decreasing $Q$.

\subsection{MHD simulations: dependence of the gravito-turbulent state on $B_{y_0}$}
\label{mhd_runs}
We first study the `ideal case' in which we do not include any
explicit resistivity. We performed a series of simulations by fixing
the cooling time $\tau_c=20\,\Omega^{-1}$ and the adiabatic index
$\gamma=2$, but varying the background toroidal field $B_{y_0}$ (or
equivalently $\beta_0$). We found that for this particular cooling
time, all simulations with $\beta_0 \gg 1$ reach a steady turbulent
state without developing massive clumps. Although we started with a
pure uniform toroidal field, the geometry of the magnetic field in the
nonlinear turbulent regime becomes very intricate and tangled. In some
cases, it is amplified and the average gas to magnetic pressure ratio
measured  in the saturated turbulent regime
\begin{equation}
\beta_{t}=\dfrac{2\left<\Sigma c_{s}^2\right>}{\left<B^2\right>},
\end{equation}
can differ greatly from the initial plasma parameter $\beta_0$. 
In the turbulent state, three
different forces emerge in the leading order balance, namely the Lorentz,
pressure gradient, and gravitational forces. By  
varying $\beta_0$, we found three different regimes 
characterised by the relative importance of the Lorentz force with respect to other forces. 

\subsubsection{First regime: $\beta_0,\,\beta_t\gg 1,\,\,E_G \gg E_m$}
\begin{figure*}
\centering
\includegraphics[width=\textwidth]{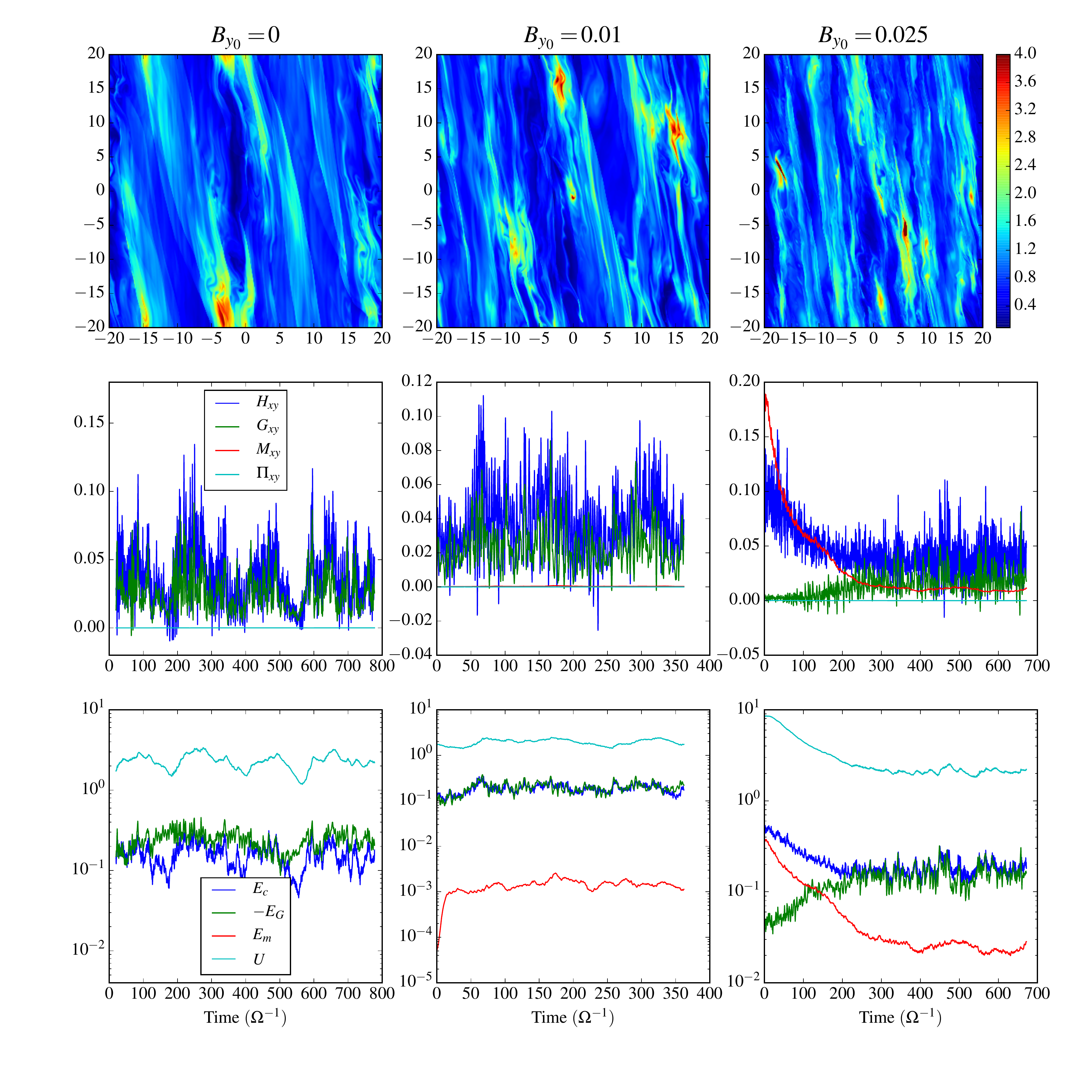}
 \caption{Simulations of gravitoturbulence in the regime $\beta_0\gg
   1,\,\,E_G\gg E_m$, computed for fixed
   $\tau_c=20\Omega^{-1}$ and $\gamma=2$.  From left to right,
   $B_{y_0}=0$, $B_{y_0}=0.01$ and $B_{y_0}=0.025$. Top panels:
   snapshots of the surface density $\Sigma$ in the disc plane
   ($x$,$y$). Centre panels: the time evolution of the Reynolds
   ($H_{xy}$), gravitational ($G_{xy}$), Maxwell ($M_{xy}$) and
   viscous ($\Pi_{xy}$) stresses. Bottom panels: the time evolution of
   kinetic $E_c$, gravitational $-E_G$, magnetic $E_m$ and internal
   energy $U$ in a  logscale. The simulation in the second column with
   $B_{y_0}=0.01$ was started from a hydrodynamic run. The third
   simulation, with $B_{y_0}=0.025$, was initiated from a MHD run with
   $B_{y_0}=0.05$, which explains why the magnetic energy and Maxwell
   stress first decrease before reaching a steady state.}
\label{fig_multi_averages1}
 \end{figure*}
The first regime corresponds to the case of a small Lorentz force 
compared to the gravitational and pressure forces. The magnetic field
is completely slaved to the gravito-turbulence and its back-reaction
on the fluid motion is insignificant or weak. The field can be stretched or
compressed so that it grows until reconnection processes take place
and destroy it. Figure \ref{fig_multi_averages1} shows three
different simulations obtained respectively for $B_{y_0}=0$,
$B_{y_0}=0.01$ and $B_{y_0}=0.025$ ($\beta_0=\infty$, $\beta_0=20000$
and $\beta_0=3000$). In the case of a weak but non-zero magnetic field,
the flow undergoes a transient phase, before it reaches a steady turbulent state. The final
state looks much like the hydrodynamic one although turbulent structures
appear on slightly smaller scales. The
center panels of Fig.~\ref{fig_multi_averages1} indicate that the
Maxwell stress is small compared to the Reynolds and gravitational
stresses. The time evolution of the energy budget is shown in the
bottom panels. It is clear that magnetic energy remains at least an
order of magnitude smaller
than the other sources and is dynamically insignificant to a first
approximation. 

Figure \ref{fig_averages_By} shows some key dimensionless quantities averaged
in space and time. The Toomre parameter $Q$ remains close to the
hydrodynamic value, although it slightly increases from $B_{y_0}=0$ to
$B_{y_0}=0.05$. This result suggests that the temperature regulation
and energy balance are only marginally affected by magnetic fields in this
regime. We note that $\beta_t$, which is  directly related to $E_m/U$, 
increases very rapidly as a function of $B_{y_0}$, 
even if it remains much smaller than the other energy ratios.

\subsubsection{Second regime: $\beta_0,\,\beta_t> 1,\,\, E_m
  > E_G$}
\label{plasmoid_regime}
\begin{figure*}
\centering
\includegraphics[width=\textwidth]{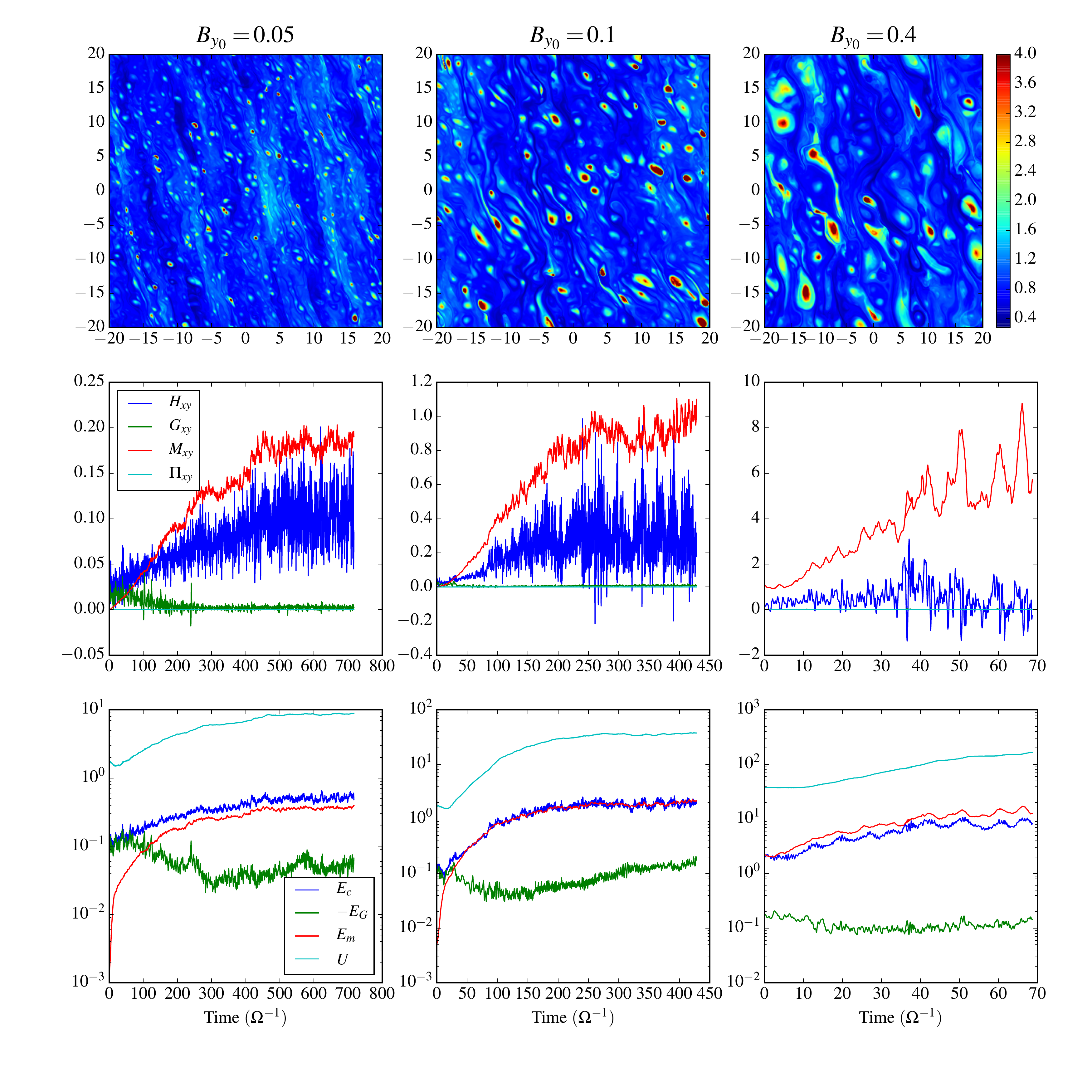}
 \caption{Simulations of gravitoturbulence in the regime $\beta_0\gg
   1,\,\,E_m > E_G$, computed for fixed
   $\tau_c=20\Omega^{-1}$ and $\gamma=2$.  From left to right,
   $B_{y_0}=0.05$, $B_{y_0}=0.1$ and $B_{y_0}=0.4$. Top panels:
   snapshots of the surface density $\Sigma$ in the disc plane
   ($x$,$y$). Centre panels: the time evolution of the Reynolds
   ($H_{xy}$), gravitational ($G_{xy}$), Maxwell ($M_{xy}$) and
   viscous ($\Pi_{xy}$) stresses. Bottom panels: the time evolution of
   kinetic $E_c$, gravitational $-E_G$, magnetic $E_m$ and internal
   energy $U$ in a logscale. The simulations shown in the first and
   second columns were started from the hydrodynamic state while the
   third was initiated from the MHD steady state
   obtained for the stronger field run $B_{y_0}=0.1$.}
\label{fig_multi_averages2}
 \end{figure*}
\begin{figure}
\centering
\includegraphics[width=\columnwidth]{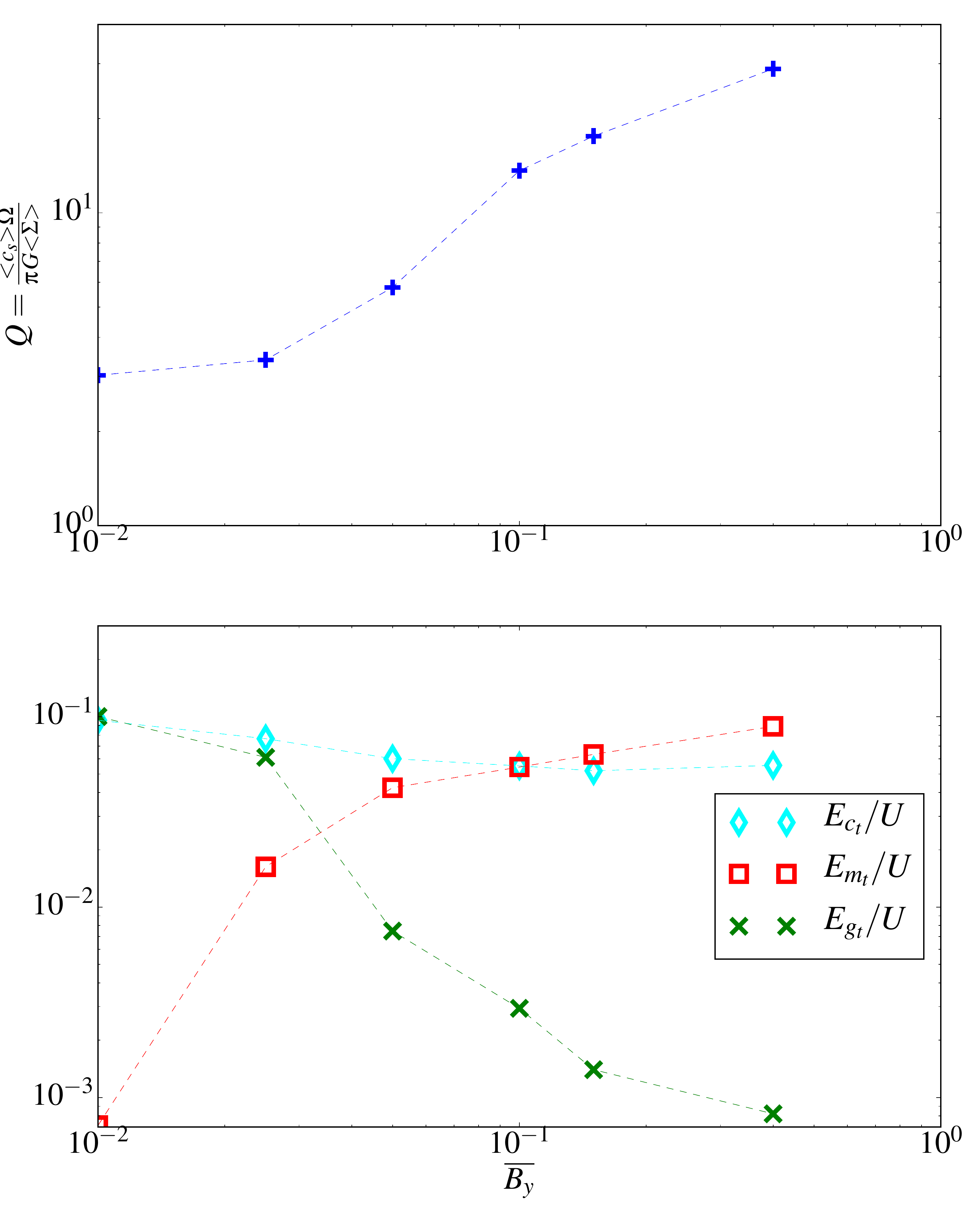}
 \caption{Top panel: the average Toomre parameter $Q$  as a function
   of $B_{y_0}$. Bottom panel: ratios of kinetic, magnetic and
   gravitational perturbation energy to the total internal energy for different $B_{y_0}$. The yellow dashed line is the turbulent angular momentum transport coefficient $\alpha$. All these simulations have been computed for a fixed $\tau_c=20\, \Omega^{-1}$ and a fixed adiabatic index $\gamma=2$.}
\label{fig_averages_By}
 \end{figure}

For stronger imposed fields we enter a second regime corresponding to
when the Lorentz force is
comparable to or larger than the gravitational force, but still smaller
than pressure gradients. Unlike the first regime, the
back reaction of the magnetic field on the fluid motion is no longer
negligible. Figure  \ref{fig_multi_averages2} shows three
simulations obtained respectively for $B_{y_0}=0.05$, $B_{y_0}=0.1$ and $B_{y_0}=0.4$ ($\beta_0=800$, $\beta_0=200$ and $\beta_0=12.5$).
The plots in the center row indicate that the Maxwell stress $M_{xy}$
is now the largest contributor to the total stress and therefore to
the angular momentum transport. Although the total stress increases with
$B_{y_0}$  owing to the new term $M_{xy}$, $\alpha$ remains constant ($\alpha=0.022$) as it only depends on the cooling time (see \citet{Gammie2001}). 
The magnetic energy is strongly enhanced in this regime and becomes
much larger than the background. Figure \ref{fig_averages_By}
shows that the ratio between magnetic energy and internal energy
increases with $B_{y_0}$  and saturates at larger $B_{y_0}$. The plasma is close to equipartition between magnetic energy and kinetic energy, as $E_c/U$ and $E_m/U$ tend to a similar value.

Figure \ref{fig_averages_By} tells us that the internal energy increases
significantly with $B_{y_0}$ because the average $Q$ and sound speed
increase by an order of magnitude between $B_{y_0}=0.05$ and
$B_{y_0}=0.4$. For $B_{y_0}=0.1$, the average temperature in the box
is multiplied by a factor ten compared to the temperature in hydrodynamic
simulations. This surprising result shows that a gravitoturbulent state can
exist at values of $Q$ much larger than in hydrodynamics.   
The greater temperatures are associated with the formation
of elongated current sheets and consequent heating at those locations
through magnetic reconnection (and associated shocks). The substantial amplification of the
background field by the turbulence and its subsequent dissipation 
provides a powerful and additional
source of heat. This is analysed in more detail in section
\ref{heating_source}.  

A consequence of this rise in temperature is
that the gravitational instability becomes weaker, as indicated by the
gravitational stress which clearly decreases by one or two orders of
magnitude. Figure \ref{fig_averages_By} shows that most of the
gravitational energy present in the small $B_{y_0}$ regime has been
replaced by magnetic energy. It is then reasonable to ask whether
self-gravity is important at all in this state. To check that, we ran a
simulation starting from developed gravitoturbulence and then switched off
the gravitational term. We found that as soon as self gravity is
suppressed, the turbulent kinetic and  magnetic energy decay to
negligible levels by $t \sim 100\, \Omega^{-1}$. This indicates that
 self-gravity,
 however weak it is, still plays a crucial role in sustaining the
 turbulence. 

The fact that graviturbulent activity persists for such
 large $Q$ may have something to do with the relaxation of
 angular momentum conservation by the magnetic stresses. It is likely that
transport via a tangle of magnetic
 fields weakens the stabilising effect of rotation, permitting GI to
 operate for larger $Q$. A similar effect is witnessed when explicit
 viscosity is included in the linear theory (e.g.\ Schmit \&
 Tcharnuter 1995) and probably in the
 nonlinear onset of GI. It should be acknowledged that in pure hydrodynamics
it is not yet understood what sets the level of the saturated $Q$;
 adding a magnetic field  must further 
complicate the
problem.

The top panels in Fig.~ \ref{fig_multi_averages2} show that the
plasma is characterized by dense clumps whose size are comparable to or
smaller than $H_0$. These `plasmoids' are associated with magnetic
island structures and evolve in a turbulent background that 
resembles the hydrodynamic state (non-axisymmetric waves
amplified transiently and dissipated into shocks). These plasmoids
appear
to resist the shear and the shocks that propagate
through them. A detailed analysis of these structures is  provided in
 section \ref{plasmoid}. Note that the magnetic islands are 
reminiscent of compressible turbulent 2D MHD simulations in which a
 forcing term is included \citep{lee03}. 

\subsubsection{Third regime: $\beta_0,\,\beta_t \leq 1,\,\,E_m \gg E_c$}

As we approach $\beta_0=1$, the fluid becomes magnetically dominated
and the gravitational term is reduced. For $\beta_0=1$  and
$\beta_0=0.5$, which correspond probably to an unrealistic regime for
astrophysical discs, our simulations show that the fluid motion is
completely frozen into the magnetic field lines and unable to move due to
strong magnetic tension. The toroidal field acts like a
`straightjacket', and steady turbulent states
cannot be achieved. Flux tubes with a coherent radial length larger
than the non-axisymmetric gravitational structures develop and
sometimes form regions of high density that collapse rapidly after
a few orbits. These  structures are very similar to those obtained by
\citet{lee03}. We note  that this regime is extremely challenging 
numerically, especially with our fine resolution, since the time step becomes very small.  

\subsection{Fragmentation criterion}
\label{fragmentation_B}
\begin{figure}
\centering
\includegraphics[width=\columnwidth]{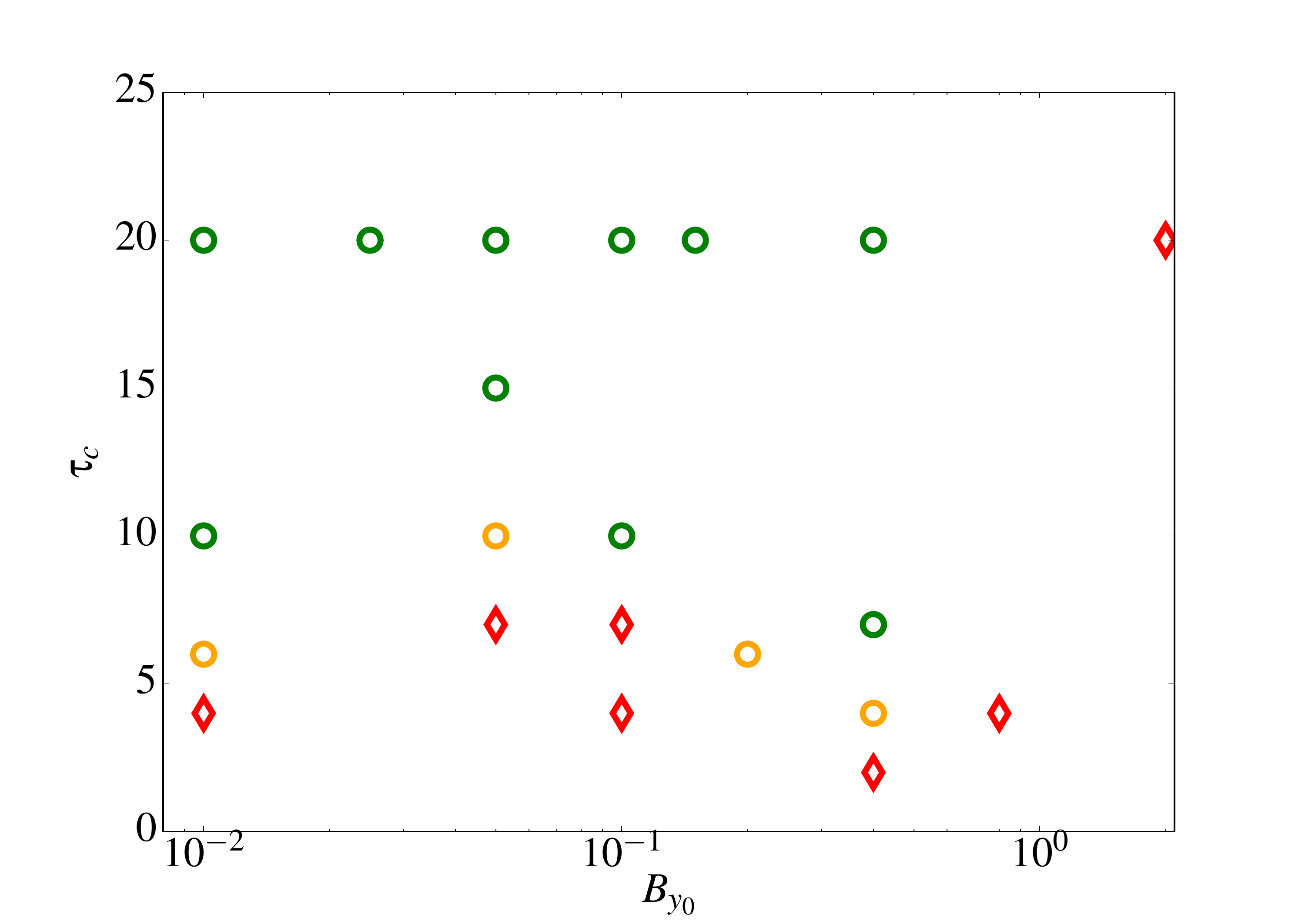}
 \caption{Fragmentation criterion as a function of the cooling time
   and $B_{y_0}$. Green circles indicate simulations that do
   not fragment at all. Orange circles indicate  simulations
   that remain turbulent but for which one or several transient
   fragments are observed. Red diamonds correspond to simulations where
   the disk collapses into one or a few massive clumps.}
\label{fig_frag_by}
 \end{figure}
In this section we vary the cooling time $\tau_c$ in order to
 study the onset of fragmentation. Each simulation is 
run for $100\Omega^{-1}$. For all values of the
background field $B_{y_0}$, there is a critical $\tau_c$ below
which
the disc collapses into massive clumps (within the simulation time). 
Figure \ref{fig_frag_by} shows how this critical $\tau_c$ varies
with $B_{y_0}$. In the first regime,
 $E_G \gg E_m$, the critical cooling time is 
multiplied by a factor $\sim 2$ between $B_{y_0}=0$ and $B_{y_0}=0.05$.
Weak magnetic tension helps fragmentation by
extracting extra angular momentum from within a potentially collapsing
region, thus thwarting the Coriolis force.

In contrast, the plasmoid-dominated regime, $E_m \gtrsim E_G $, witnesses
an unexpected decrease in the critical cooling time with $B_{y_0}$.
This is probably due to
the enhanced heating generated in
the presence of a stronger imposed field. An increased
magnetic field causes the gas to be signifcantly hotter, as explored
in the previous subsection. 
The associated pressure prevents fragmentation and
overwhelms the direct destabilising effect of the magnetic field in collapsing 
a plasmoid, via magnetic tension and pressure.

Overall, the relative variation of the critical
cooling time with $B_{y_0}$ remains small, which suggests that
magnetic fields, despite their strong influence on the turbulent
properties, do not dramatically change the fragmentation
criterion. Note that, similarly to the hydrodynamical case, the average
$Q$ decreases when $\tau_c$ is decreased ($Q\sim 30$ for $\tau_c=20$
and $Q\sim 18$ for $\tau_c=7 \,\Omega^{-1}$ in the case of
$B_{y_0}=0.4$). No criterion for fragmentation depending on
$\alpha$ or $Q$ can be obtained simply in that case. 

\section{Effects of resistivity}
\label{resistive_runs}

Up to now we have explored only `ideal' MHD --- the particulars of the grid 
have been taking care of the reconnection, diffusion, and
thermalisation of magnetic field. In this section we include magnetic resistivity
explicitly to better control this process and also to push our models
to regimes relevant to the more resistive radii in protoplanetary
disks. We find that increasing diffusion, unsurprisingly, impedes the
build up of the strong fields witnessed in section \ref{mhd_runs}; the
field slips through the
turbulent gas and is no longer wound up, stretched, and amplified
as efficiently. 

In protoplanetary discs, the magnetic Reynolds number Rm
 is directly proportional to the gas's ionisation
 fraction, which is determined by interparticle collisions, 
cosmic rays and X-rays, radioisotopes,
molecular recombination, and dust grain physics
\citep{armitage11}. In a minimum mass solar nebula (MMSN) model, 
estimates for the midplane Rm vary from $\sim 0.1$
at 5 AU to $10^3$ at 10 AU, to greater than $10^4$ at larger radii \citep{simon15}. 
In fact, both ambipolar diffusion and the Hall effect are more important 
than Ohmic diffusion at the latter two radii. The equivalent ambipolar
magnetic Reynolds number is defined as $R_A =
\Omega H^2 \nu_{in} \rho^2 x_e/ B^2$ where $\nu_{in}$ is the ion-neutral
collision rate and $x_e$ is the ionisation fraction. This number
varies roughly between 0.1 to 10 times $\beta$ 
between 5 and 100 AU \citep[e.g.][]{simon15}. If we are permitted to 
crudely model ambipolar diffusion by Ohmic diffusion in our
simulations, 
then our effective Rm 
should take values between 1 and $10^4$.

The reader should be aware, however, that 
the above estimates for non-ideal MHD were derived with the MMSN model,
which best describes an older type-II disk, which is insufficiently
massive to suffer GI. The relative strengths of Ohmic and ambipolar
diffusion will differ in a GI-unstable type-0 system, which will be denser
(hence the ions and neutrals better coupled) but also less
well-ionised because optically thicker. The above estimates hence only
serve as a rough guide, to fix ideas.   

An additional
complicating factor is that the ionisation fraction (and hence Rm and $R_A$) 
depends on height, and so the gas at 
different vertical levels is not coupled to the magnetic field in the
same way. 
These issues make it less than straightforward to assign a simple
Ohmic diffusivity to 2D simulations, and in fact to interpret the
role of diffusion in two dimensions generally.

One other extremely important
ingredient, neglected in our work, is the MRI. Though absent in the
heart of dead zones $\sim 5$ AU, it may appear in the more favourable
ionisation conditions at larger radii, though the details of its
prevalence are exceptionally complicated and the subject of
intensive research. The outcome depends not only on the
(poorly constrained) ionisation profile, but on the orientation,
strength, and
existence of a net magnetic field, with simulations showing that the midplane
can be completely laminar, undergo bursts of turbulence, or sustain a sluggish    
form of the MRI. The
surface regions, on the other hand, may launch a magnetocentrifugal 
wind or suffer vigorous turbulence
\citep{simon13b,simon15}.  
We cannot hope to adequately model this 
physics in our 2D simulations, but hope that its diffusive aspects can
be roughly described by a constant resistivity.

\begin{table}     
\centering 
$B_{y_0}=0.05$ i.e $\beta_0=800$
\begin{tabular}{c c c c c c}          
\hline                        
Rm & $Q$ & $E_m$ & $E_m/U$ & Plasmoids \\    
\hline  
	0 (hydro) 	& 3.02 & x & x & x \\
	
	10	& 3.16 & 0.0011 & 0.002 & NO \\                             
    100 &  3.2 & 0.005 & 0.0026 & NO\\      
    500 & 3.4 &  0.025  & 0.012 & NO\\
    1000 & 4.9 & 0.16  & 0.031 & YES (very few) \\
    5000 & 5.7 & 0.27   & 0.037 & YES\\
    ideal approx. & 5.8 & 0.36 & 0.042 & YES \\
    
\hline  
\end{tabular}                                         
\vspace{1cm}

\centering   
$B_{y_0}=0.1$ i.e $\beta_0=200$
\begin{tabular}{c c c c c c}          
\hline                        
Rm& $Q$ & $E_m$ & $E_m/U$ & Plasmoids \\    
\hline  
	0 (hydro) &3.02 & x & x & x \\
    100 & 3.1 & 0.024 & 0.012 & NO\\      
    250 & 7.6 & 0.28 & 0.026 & YES (very few)\\
    500 & 10.9 &  0.84    & 0.035 & YES\\
    1000 & 12.5 & 1.35  & 0.045 & YES \\
    5000 (hlld) & 13.6 & 1.78 & 0.05 & YES\\
    ideal approx.& 13.7 & 1.88 & 0.053 & YES \\
    
\hline 
\end{tabular}
\vspace{0.5cm}
\caption{Average $Q$, magnetic energy, and the ratio of magnetic to internal energies
  for different Rm. The first table
  corresponds to  simulations with $B_{y_0}=0.05$, and the second to
  $B_{y_0}=0.1$. The last column indicates whether plasmoids appear. 
  All simulations were conducted
  with the hll solver except for $\text{Rm}=5000$ and
  $B_{y_0}=0.1$ which was undertaken with the hlld
  solver.} \label{table1}
\end{table}

\subsection{Resistive turbulent simulations}

 To determine the effect
of Ohmic diffusion on MHD gravito-turbulent states, we performed a
series of simulations with explicit resistivity by taking a fixed
$B_{y_0}$ and varying the magnetic Reynolds number. We scanned a large
range of Rm, straddling midplane values typical of dead zones and larger
radii. Our simulations were initiated from a
saturated state computed in the ideal limit and run until a new steady
state  was found. 

Table \ref{table1}a) sums up the different results obtained for
$B_{y_0}=0.05$ (equivalently $\beta_0=800$). 
As expected, when $\text{Rm}= 5000$ the turbulent
state 
differs little from the `ideal case' because the numerical and physical Rm
are of the same order. For such values of Rm, Ohmic diffusion is probably
unresolved. 
However, the statistical properties of the
turbulence change quite drastically when $\text{Rm}\lesssim 1000$. For
this transitional Rm, the magnetic energy is roughly halved
while $Q$ drops to 4.9 (from 5.8). Another change is
that the number of plasmoids in the box is considerably reduced while
their typical density decreases by a factor 4.  Below
$\text{Rm}=500-1000$, plasmoids structures disappear and $Q$ approaches
its hydrodynamical value. At these Rm, diffusion impedes the
build up of large magnetic energies (that may be subsequently
thermalised) and the disk is hence cooler.
Results for a stronger imposed field $B_{y_0}=0.1$ ($\beta_0=200$)
are presented in Table \ref{table1}b). Now the magnetic energy
and $Q$ decrease sharply around a lower critical $\text{Rm}\sim
100-250$, and plasmoid structures disappear for
$\text{Rm}\lesssim 250$.

To conclude, the main effect of resistivity is to reduce the average
turbulent magnetic energy stored in the fluid. As a consequence, there
is less free energy to be dissipated into heat and the mean
temperature decreases. Below some critical $\text{Rm}_c$ that seems to
scale as $\beta_0$, the turbulence becomes decoupled from the
magnetic field. This result suggests that a key quantity to study the
transition between `quasi-hydrodynamical' and plasmoid-dominated  
MHD turbulence might be the Elssaser number $\Lambda \sim \text{Rm}\,\beta_0^{-1}$.  

\subsection{Zero net flux simulations and decay rate}

As mentioned in section \ref{ic}, a 2D turbulent flow cannot sustain
a dynamo field, which means that if we start with a zero net toroidal
flux, the magnetic field is expected to decay within a finite
time. However, the decay time can be exceptionally long, due to
compressibility, the geometry of the initial
field, and a large Rm (see also Ivers \& James 1984). It may even be
possible to sustain a magneto-turbulent state throughout a great
fraction of a disk's life. To give an
estimate of the decay timescale, we performed two different
simulations with zero net flux. In the first one, labelled `Zs', we
started from a state computed with $B_{y_0}=0.1$, removed the mean
component of the toroidal field and then let the flow evolve in time. In
the second one, labelled `Zl', we started from the same state but we
added at $t=0$ a sinusoidal $B_{y_0} \propto \sin(2\pi/L_y)$ with an
energy equivalent to the one with uniform background field. Both
simulations were performed with explicit resistivity and
$\text{Rm}=5000$. In the Zs case, the magnetic energy $E_m$ decays
to negligible values
by 150 $\Omega^{-1}$, which corresponds roughly to the decay time
expected. Indeed if the turbulent magnetic structures are of scale
$\sim H_0$, then the resistive decay time is given by
$H_0\,\text{Rm}/4\pi^2 \sim 126 \, \Omega^{-1}$. In the Zl case, we
found however that the initial magnetic field is retained over
at least $ 1000\,
\Omega^{-1}$  while  magnetic energy stays virtually constant
throughout the
simulation. This is expected because the estimated decay time
for the large-scale $B_{y_0}$ is of the order $2\times10^5\, \Omega
^{-1}$, comparable or longer than the  disc viscous
timescale. 

\section{Current sheets and plasmoids}

\subsection{Heat sources and currents sheets}
\label{heating_source}

We showed in section \ref{mhd_runs} that for intermediate $B_{y_0}$,
the Maxwell stress produces a large contribution to the total stress and
provides an additional source of thermal energy. The build up of
magnetic energy is another source, once it is dissipated
via current sheets or related
structures \citep{parker72,cowley97}. 
As the heat generated by magnetic fields
drastically alters the thermodynamic state  
of the turbulence, and in particular the average $Q$, it is crucial to
better understand it.

\subsubsection{Mean features}

To identify the main source of heat in our MHD simulations,  
we investigated the relative importance of each term in the averaged equation for internal energy:
\begin{equation}
\label{eq_average_heat}
\dfrac{d \langle U \rangle}{dt}= \langle - P  \nabla \cdot \mathbf{u}
\rangle +\langle D_{\nu}+D_{\eta} \rangle -\langle U \rangle/\tau_c.
\end{equation}
Physically, the heat can be generated through two different processes:  reversible compression or expansion of the gas 
which is associated with the term
\begin{equation}
W_{PV}=- P  \nabla \cdot \mathbf{u},
\end{equation}
and irreversible dissipation like viscous and Ohmic friction, whose dissipation rates are respectively:
\begin{equation}
D_\nu= \boldsymbol{\Pi} : 
 \mathbf{\nabla v} \quad \text{and} \quad D_\eta=\eta \left(\nabla \times \mathbf{B}\right)^2.
\end{equation}
The main difference between these sources is that pressure work
 $W_{PV}$ (also called `pressure-dilatation') can have either a positive
or negative sign, meaning that the energy transfer between kinetic
and thermal modes can be in either direction. In contrast, irreversible
processes transfer energy from the kinetic to thermal channels only. 

We
analysed the net heat budget for $B_{y_0}=0.1$ and Rm $=500$ by
averaging eq.~(\ref{eq_average_heat}) in time over $150 \Omega
^{-1}$. We found that almost 45\% of the budget is represented by pressure
work $W_{PV}$, 20\% by the Ohmic term $D_{\eta}$ and 6\% by the
viscous term $D_{\nu}$. The leftover is taken up by numerical
dissipation. Note that when a HLLD solver is used, the viscous term
becomes 11\% but the average $Q$ and temperature are unchanged. The
rather significant amount of numerical dissipation is not
surprising and arises because of the presence of thin shocks layers which are
difficult to resolve viscously. However, as discussed in section 2,
only a tiny fraction of energy is lost, the numerical dissipation is
mostly recycled as heat in a way approximating real
microscopic dissipation inside a
shock. 

A notable result is that a large fraction of the heat comes from the
reversible expansion of the gas through the pressure dilatation term
$W_{PV}$, which oscillates between positive and negative values, with a
frequency $\approx\Omega$ but which is positive on average. 
In most studies, this reversible
heating is considered irrelevant because a fluid parcel in the disc
is thought to relax adiabatically and return to its unperturbed state shortly
after the passage of a spiral wave or a shock
\citep[e.g.,][]{rafikov16}. On
average, the heat generated through an expansion is removed by a
subsequent relaxation because there is a similar degree
 of compression and
expansion in the gas ($\langle \nabla\cdot \mathbf{u}
\rangle=0$). This is true when the gas
can relax adiabatically on a timescale
much shorter than the cooling time, and when a gas parcel encounters
waves or shocks on a timescale
longer than the adiabatic relaxation time. 
In our simulations neither need be the case. The fluid
endures a turbulent forcing so strong that it has no
time to relax between each compression or shock crossing. A similar
net reversible compressible energy transfer, due to pressure
dilatation, has been observed in 2D and 3D hypersonic
compressible turbulence, with no radiative cooling, in a
separate non-disk context \citep{zeman91,sarkar92}. In particular,
for shear flows, this transfer can be comparable to the compressible
viscous dissipation term and contributes to a reduced growth of
turbulent kinetic energy when the  
flow is integrated over a long time (the missing part being transferred to internal energy).\\

\begin{figure}
\centering
\includegraphics[width=\columnwidth]{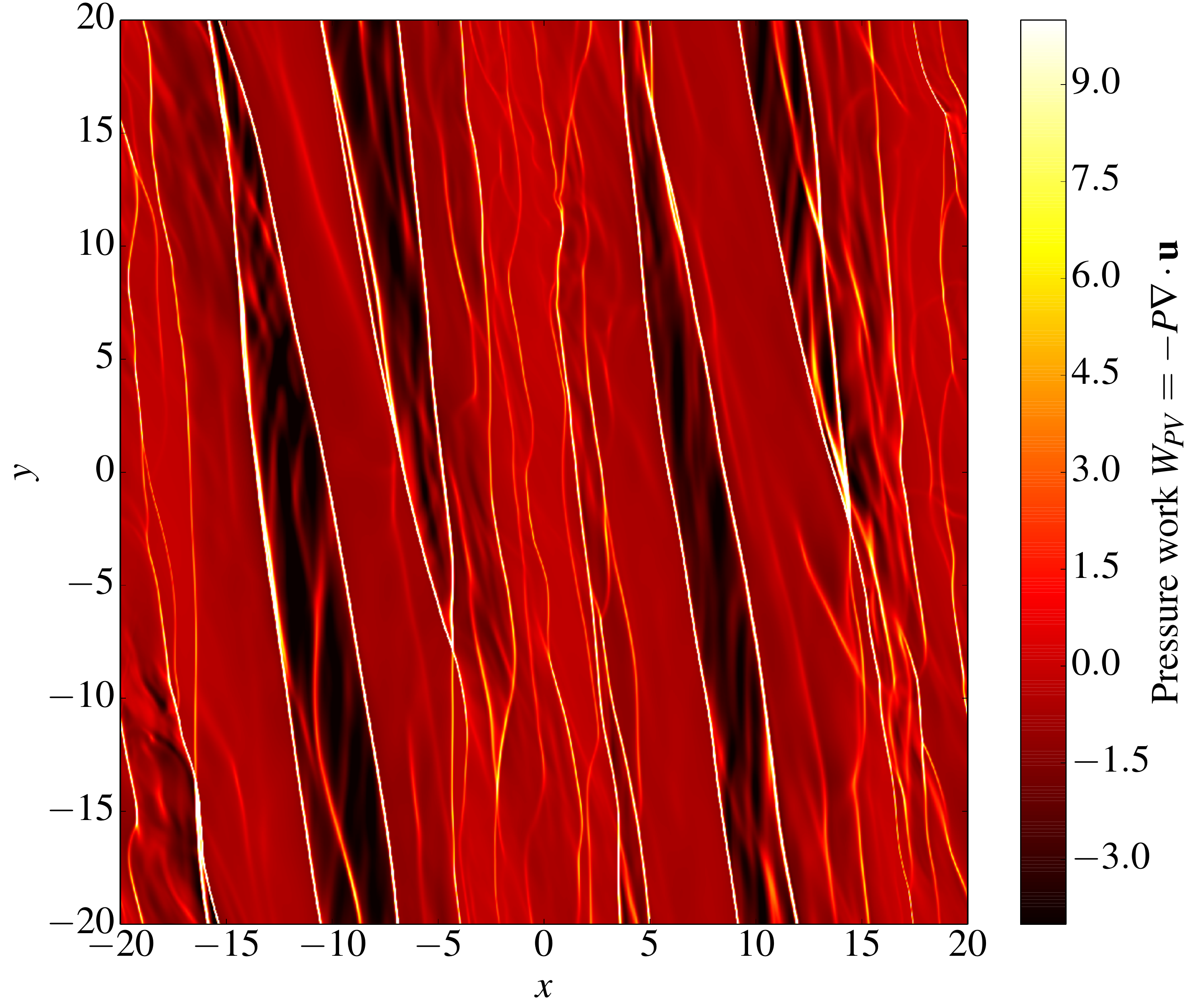}
 \caption{A snapshot of the distribution of pressure dilatation
   $W_{PV}$ in hydrodynamic gravitoturbulence ($B_{y_0}=0$). 
  Bright and white colors indicates heating by compression, 
  dark and black indicates region where the gas expands and relaxes. 
 The intensity has been intentionally saturated at $W_{PV}=10$ 
 but the real maximum is  $W_{{PV}_{max}}\sim 150$.}
\label{fig_heat_source}
 \end{figure}

\begin{figure}
\centering
\includegraphics[width=\columnwidth]{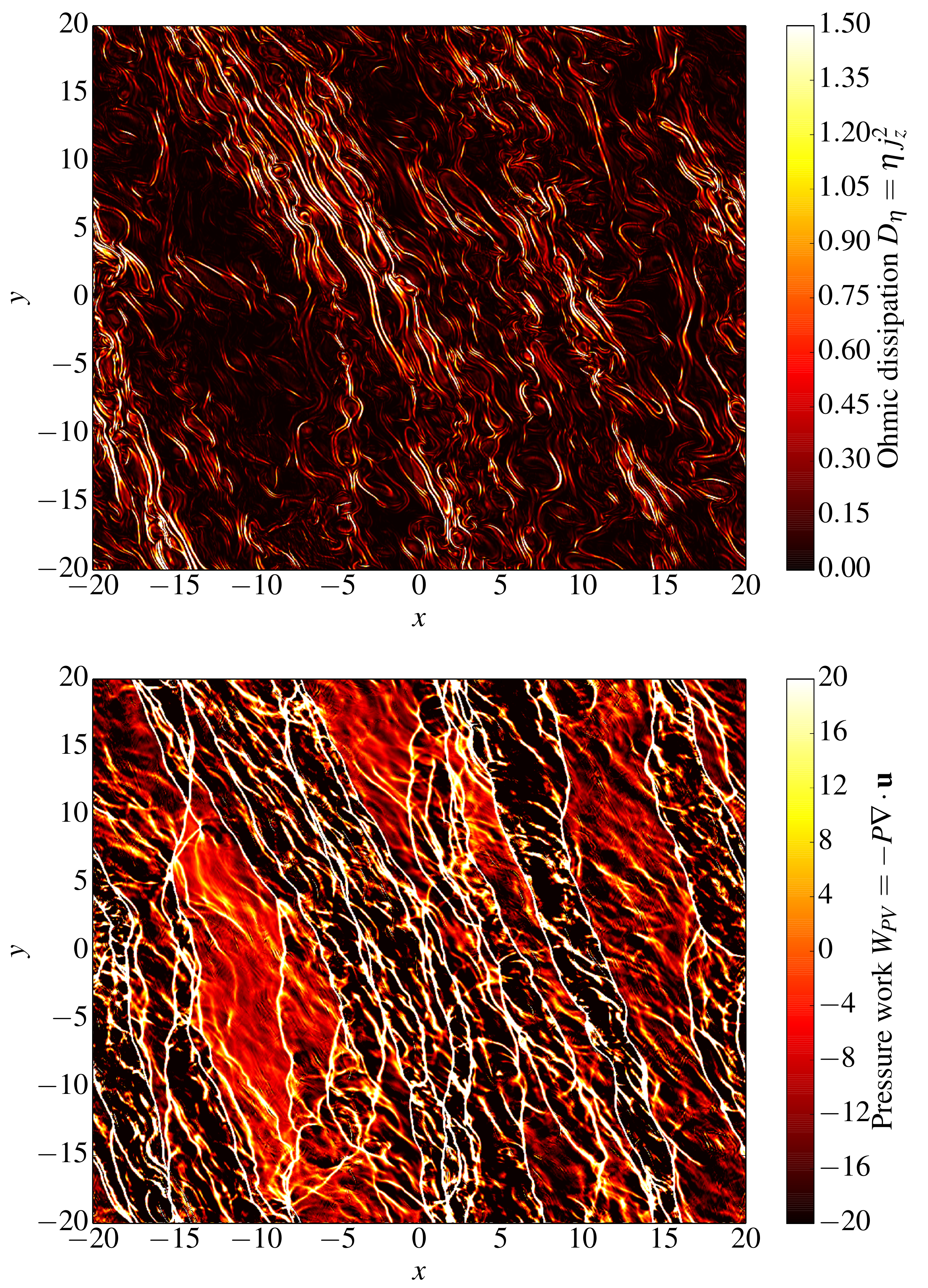}
 \caption{Heat sources in a quasi-steady turbulent state obtained for
   $B_{y_0}=0.1$ and $\text{Rm}=500$. The top panel represents the
   Ohmic dissipation $D_\eta$ whereas the bottom panel represents the
   pressure dilatation $W_{PV}$.  
   The intensity has been 
  intentionally saturated at $W_{PV}=20$ but the real maximum is 
   $W_{{PV}_{max}}\sim 5000$, much greater than in hydrodynamical runs.}
\label{fig_heat_source2}
 \end{figure}

\subsubsection{Dissipative structures}

In addition to the average budget for the internal energy, we analysed
the spatial distribution of the heat sources in the hydrodynamic case
and in a magnetized gravito-turbulent flow with $B_{y_0}=0.1$. Figure
\ref{fig_heat_source} shows that when $B_{y_0}=0$, the main sources
(here the reversible part) are located in  very thin  azimuthally
elongated structures. These thin layers correspond to shock waves
that propagate within the fluid and are associated with the nonlinear
evolution of large-scale gravitational wakes. The
black/dark regions correspond to expanding gas ($\nabla\cdot
\mathbf{u}>0$) where the pressure is found to be maximum. 

Figure \ref{fig_heat_source2} presents a snapshot from a magnetic
simulation. The second panel shows again that the pressure
work is concentrated into very thin filaments, which reveal the
location of shocks, but their geometry is tremendously more
complicated and their number clearly increased. In comparison with the
hydrodynanical case, the surface covered by these heat sources is
multiplied by a factor $\sim 10$, for $B_{y_0}=0.1$ (this is
estimated by computing the surfaces where $W_{PV}>10$). These
intricate patterns reveal also that the heat transfer is concentrated
 in smaller
 scale structures, with a typical length that seems correlated 
to the size of the magnetic field bundles.

 Although we showed
that Ohmic dissipation is not the dominant source of heat directly,
its associated current
sheets might play a crucial indirect role by generating shocks.
 Figure ~\ref{fig_heat_source2}a) shows the regions where
magnetic energy is dissipated into heat via Ohmic dissipation. These
regions clearly take the form of filamentary structures
with a small azimuthal extent (compared to $L_y$). There appears to be
a correlation between the number of such sheets and the number of shocks for which $W_{PV}$ is positive. In addition, the location of these structures seems to be found in regions of high pressure, where $\nabla\cdot
\mathbf{u}$ is actually a minimum.  These regions correspond 
to 
the self-gravity wakes that take the form of large scale non-axisymmetric bands. \\

Magnetic reconnection in current sheets is known to accelerate the
gas, producing sometimes a pair of slow-mode shocks extending outwards
from the central sheet
\citep{petscheck64,priest86,birn07,hillier16}. These shocks are known
to  be highly effective at
heating the surrounding medium. In fact, some numerical studies
 indicate that the slow mode shocks are the
primary heating mechanism in the solar corona \citep{bareford15}. In some
circumstances the energy released from these shocks can be more
important than Ohmic dissipation, as seems to be the case here. 
Though we do not go into a detailed analysis in this paper, it seems
plausible that the enhanced heating witnessed by magnetic
gravitoturbulence is caused by Ohmic reconnection in current sheets and in
the shocks generated by such reconnections.

\subsection{Plasmoids}
\label{plasmoid}
As pointed out in section \ref{plasmoid_regime}, when magnetic
fields have a moderate amplitude and Rm is not too small, the
turbulent flow displays coherent plasmoid structures and magnetic
islands. These patterns have been studied exhaustively in the
literature of magnetic reconnection
\citep{park84,ugai95,loureiro05,huang13,loureiro16} and 2D compressible
MHD turbulence \citep{lee03} but have not featured especially
in simulations of accretion disc dynamics. In this context they
deserve further attention as it is tempting to associate them with 
planet formation, possibly as sites in which dust may accumulate.
Separately, an examination of their intrinsic balances and structure
may help unveil the role of the Lorentz force 
in magnetized shear flows generally. 

\begin{figure}
\centering
\includegraphics[width=\columnwidth]{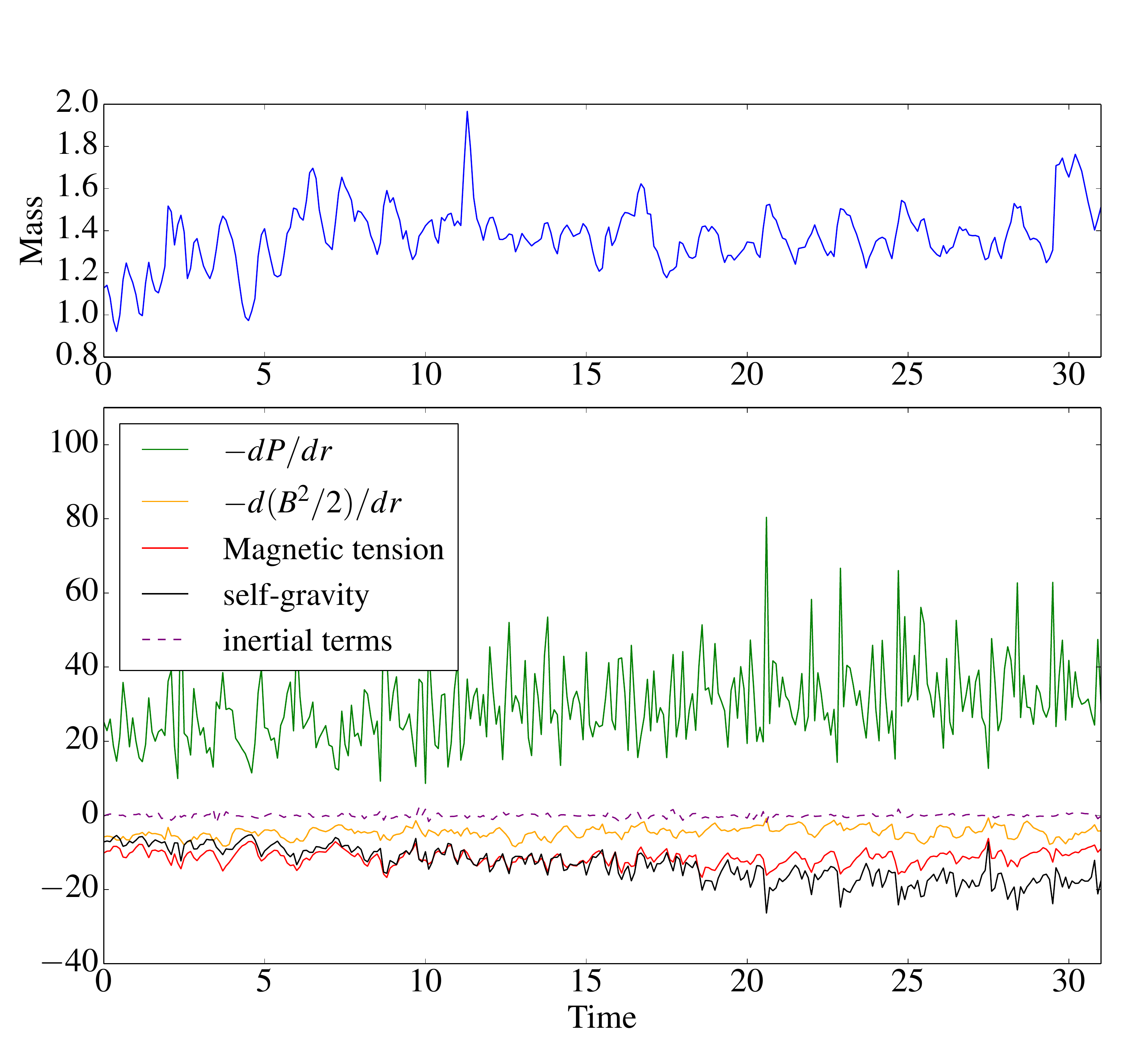}
 \caption{Top panel: evolution of the total mass in a plasmoid
   computed from a simulation with $B_{y_0}=0.05$ and no explicit
   resistivity. Bottom panel: radial forces integrated over the
   interior of
   the plasmoid in a frame  of reference centered at the pressure
   maximum. 
 The typical radius of the plasmoid (taken as our integral bound for averaging) is $r_p=0.4 H$.}
\label{fig_plasmo_eq}
 \end{figure}

\subsubsection{Are they fragments?}

The first question is what relationship these magnetic islands have
with respect to the gravitationally bound fragments that appear in
hydro simulations of GI. Figure~\ref{fig_plasmo_eq} (top) 
shows that, for $B_{y_0}=0.05$
and $\tau_c=20/\Omega$, the total mass integrated inside one of the
plasmoids does not increase with time and keeps a fixed value during
more than 30 $\Omega^{-1}$. Actually, we checked visually that they
stay stable over a much longer time. Although they form dense
structures with $\Sigma$ that can exceed 50 times the
background $\Sigma_0$, they do not  seem to be regions where the gas
is collapsing, at least for sufficiently large $\tau_c$. We conclude
that they are
resolved quasi-steady objects quite different to the
fragments seen in hydrodynamical simulations of gravitational collapse.

\subsubsection{Origins}
\begin{figure*}
\centering
\includegraphics[width=0.6\textwidth]{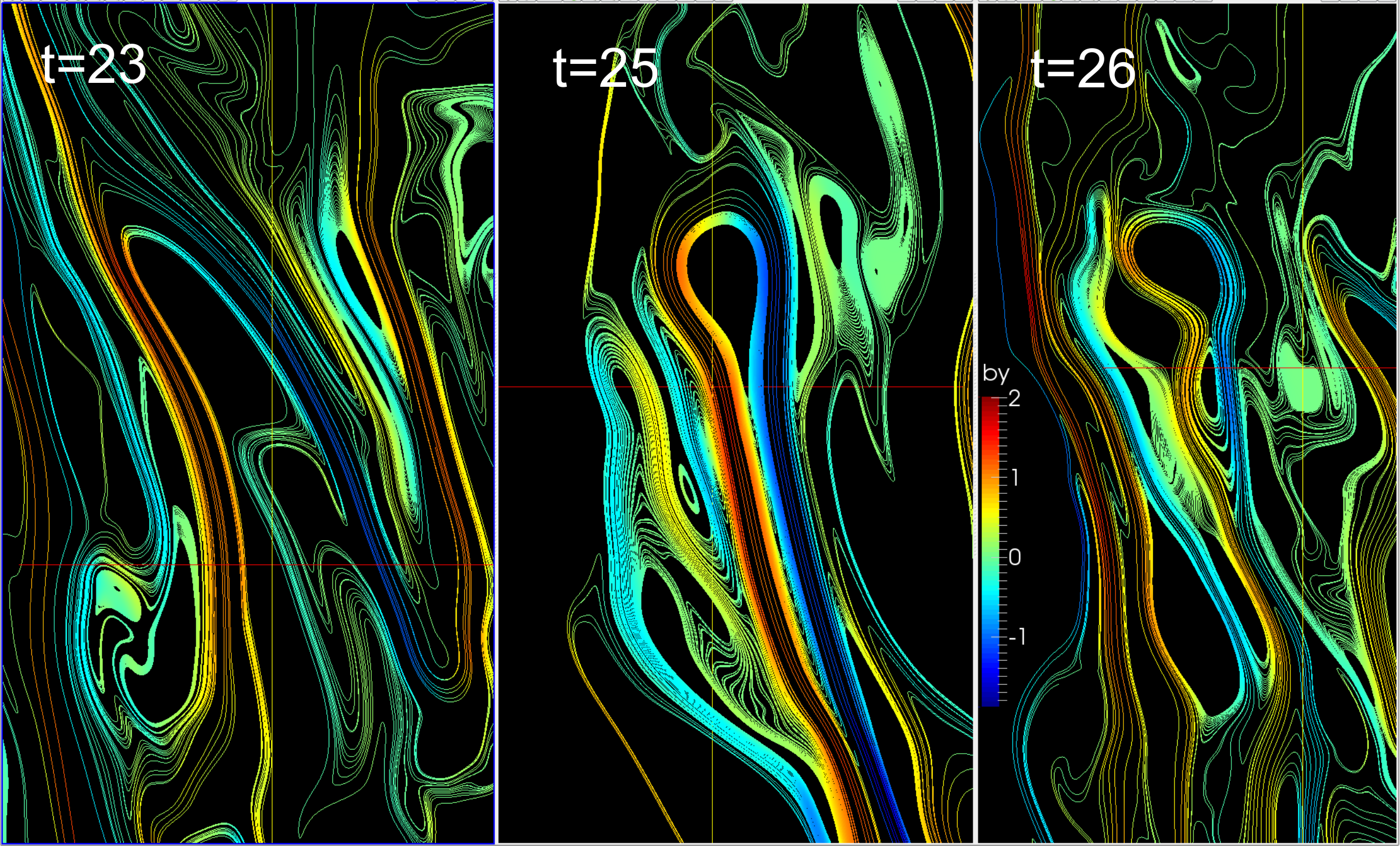}
 \caption{Three snapshots showing the evolution of the magnetic field line topology in a small  
patch centred around a region of plasmoid formation over a period of
$3\,\Omega^{-1}$. 
  Blue represents negative field polarity, while represents positive
  $B_y$. The intensity of the colour quantifies the magnitude of the
  field. The $x$-extent of the patch is $3H_0$, while the $y$-extent
  is $5H_0$.}
\label{fig_reconection}
 \end{figure*}

In order to understand how plasmoids form, we investigated the early stages
of a simulation in which magnetic islands appear. We found that this stage
occurs just after the onset of the turbulence. Figure
\ref{fig_reconection} shows the field line configuration near a
plasmoid forming region between $t=23\,\Omega^{-1}$ and
$t=26\,\Omega^{-1}$ for $B_{y_0}=0.1$. Initially straight azimuthal
field lines are stretched, folded and amplified by the turbulent
eddies. Strong positive (red) and negative (blue) toroidal magnetic
fields are then brought together. At $t=25\,\Omega^{-1}$, a current
sheet is forming as soon as the magnetic loop is closed. At
$t=26\,\Omega^{-1}$, the field lines become
possibly unstable to the tearing instability
\citep{biskamp86,loureiro05} and reconnect,
forming two magnetic islands. These snapshots (and many others like
them) suggest that plasmoids
are generated through a common physical mechanism and are not produced
artificially by the code. 

\subsubsection{Equilibrium and structure}
\label{plasmoid_structure}

\begin{figure}
\centering
\includegraphics[width=\columnwidth]{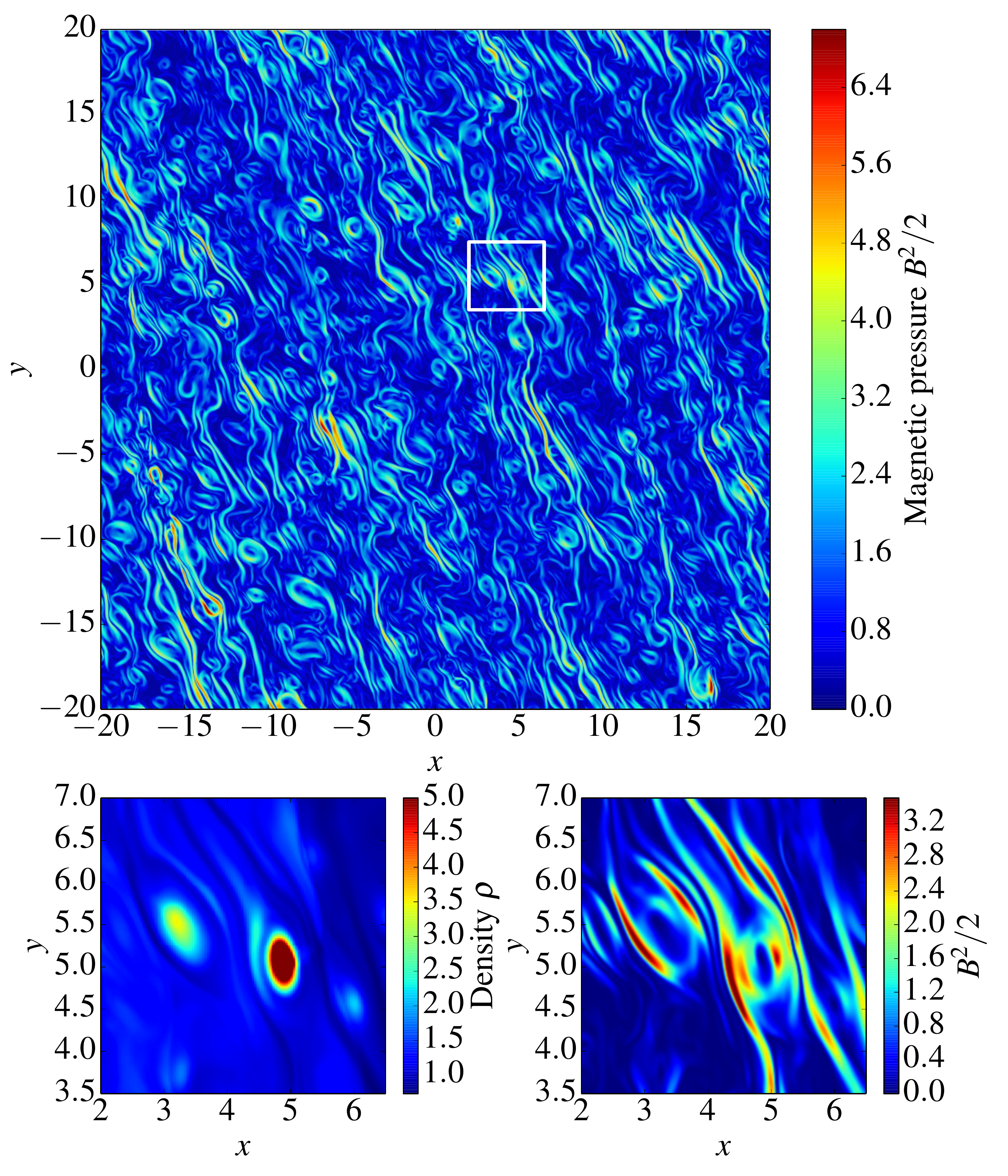}
 \caption{Top panel: total magnetic pressure  in a quasi-steady
   turbulent state obtained for $B_{y_0}=0.05$ and no explicit
   resistivity. The bottom left and right panels are respectively the
   density and magnetic pressure in the white rectangle appearing in
   the top panel.}
\label{fig_plasmo_magp}
 \end{figure}

Figure \ref{fig_plasmo_magp} shows a snapshot of a simulation computed
for $B_{y_0}=0.05$ with no explicit resistivity, containing a number of
plasmoids. Magnetic pressure forms strong ring-shape structures
surrounding each of these plasmoids that prevent the external gas from
penetrating within. To a first approximation 
 they behave as steady rigid bodies in a
sheared turbulent background. Some of them are spinning  with a net
negative vorticity (in the same direction as the shear) but their
velocity profile can be quite intricate inside.  

Gas pressure and
density are always maximum at their centres and decay
quasi-exponentially with distance to this axis. Plasmoids are cold
and their temperature is minimum at their centres. In order to
better understand their internal structure, we
computed their force balance. We chose one of our
simulation with $B_{y_0}=0.05$ and tracked a number of plasmoids by
calculating their position and velocity at each output time. Forces
are then computed in a particular frame of reference, centred upon a
given plasmoid (where the density is maximum). We used polar
coordinates around this origin so that $r$ denotes the distance to the
center of a plasmoid and $\theta$ the angle with the $x$ axis. Each
force is averaged inside the plasmoid by integrating in $r$ and
$\theta$. We defined the radial extent of a plasmoid as the radius
at which the density has dropped by  a factor 2 from its center. 

We identified one big plasmoid where the  density contrast between the
background turbulent flow and the center is
$\Delta=0.13$. Fig.~\ref{fig_plasmo_eq} shows the radial force
balance in the frame of this structure. Inside the plasmoid, the fluid
is in equilibrium between the pressure gradient (which is positive and
resisting the collapse) and all other forces (that are negative
and tend to make the gas collapse). The latter are all
of the same order of magnitude although magnetic pressure is roughly
half the magnetic tension and self-gravity. Note that inertial forces
(Coriolis and nonlinear advection) affect the
equilibrium only weakly, indicating that the pressure maxima is not
maintained by vortical motions. The plasmoids hence should not be
regarded as vortices. (We did not plot the viscous force
as it is completely negligible.) Finally, we checked that this force 
 balance is similar for several other plasmoids.

Figure \ref{fig_multi_averages2} indicates that the plasmoids sizes
increases with $B_{y_0}$. On one hand, this behaviour might be
surprising as we explained that magnetic forces seem to push the gas
inward and force the structure to contract. But on the other hand, for
the same reason as explained in Section \ref{fragmentation_B}, the
pressure increases very rapidly with $B_{y_0}$, due to the heat
generated by magnetic fields in this regime. In addition, the
gravitational force becomes less important in comparison with pressure
forces as $Q$ is increased. Therefore, a balance is still possible and
the size of the structures can even grow with $B_{y_0}$.

\section{Discussion and Conclusion}

In summary, we performed 2D shearing sheet simulations of
gravitoturbulence in magnetised accretion disks
penetrated by a net toroidal field. For moderate plasma beta and
magnetic Reynolds number, this field was twisted, warped, and greatly
amplified by the turbulent velocity fluctuations. 
 Once a quasi-steady
state was achieved the final magnetic energy could, in fact, be equal to the
turbulent kinetic energy. Once thermalised,
this additional reservoir of energy leads to a dramatic
heating of the gas, and enhanced quasi-equilibrium temperatures (and thus
Toomre $Q$'s). For example, when $\beta_0\sim 100$ and Rm$>1000$, the
mean $Q$
is amplified over the hydrodynamic value by a factor 4. For the same
$\beta_0$ but a larger Ohmic resistivity, Rm$=250$, the amplification
is a factor 2. The system can thus achieve a marginal gravitoturbulent state
in which $Q\sim 20$, and the gravitational potential energy
subdominant (though absolutely necessary for the subsistence of the steady state). For lower Rm or weaker imposed
fields these striking effects subside and the system begins to resemble the
hydrodynamical regime. We tentatively attribute the persistence of GI
activity
at such high $Q$ to the breaking of angular momentum conservation by
the tangled magnetic field. The suppression of this stabilising effect
exacerbates the GI and extends the range of gravitoturbulent activity
to hot states where it would ordinarily be stable.

The thermalisation of the magnetic energy is undertaken through the
action of small-scale current sheets, and especially the slow shocks
generated by reconnection in the sheets. The resulting heating is
highly inhomogeneous and localised in an intricate network of shock
layers. The temperature fluctuations in this network may greatly
exceed the mean temperature of the disk, and may have some
consequences for chemistry and the processing of solids (see for
example, \citet{godard09} or \cite{mcnally14}). For sufficiently
large Rm, reconnection also
generates plasmoids, long-lived magnetic islands distinct
from both vortices and gravitationally collapsing blobs. If
shown to be prevalent and robust, these structures could be of
interest to planet formation theories. 

Finally, we checked to see if magnetic fields had any impact on the
fragmentation criterion. By varying the cooling time, for different
imposed fields, we obtain critical $\tau_c$ below which the gas
fragments. In general, these critical values are not very different to
the hydrodynamical ones. Given the numerically dependent and
stochastic nature of fragmentation, it is difficult to set much store
on these results --- though the basic idea (that magnetic fields are
not so important) may be robust.

Our 2D ideal and resistive simulations are potentially relevant for
protostellar disk regions that are magnetically active but MRI stable.
As shown in vertically stratified simulations with the full
panoply of non-ideal MHD, such regions may span a significant range of
outer radii. Strong horizontal fields may be generated by the Hall effect,
and winds launched at the disk surfaces. Our numerical set-up does not
correctly capture these non-ideal effects, but nonetheless some of the
behaviour we witnessed might cross over. An additional uncertainty, in
any case, is the correct non-ideal regime for the outer radii of
gravitationally unstable class-0 disks. Previous work, and our
estimates, have been based
on the less massive MMSN model.

Our simulations may also be relevant for massive deadzones in older
disks, at the onset of GI-instigated outbursts. The newly GI active
region, if supplied by sufficiently strong magnetic fields
 (by advection from larger radii or
locally by Hall currents) could more effectively heat the gas, as
described above and thus more easily kickstart the classical MRI, as
required by certain outburst models. That said, this enhanced heating
requires somewhat larger Rm than typically supported by dead zones, and may only
be effective at the outer edge of the zone.

A final application of these results may be to the outer part of AGN discs
  which are likely to be gravitationally unstable \citep{paczynsky78},
  and susceptible to fragmentation (Goodman 2003, Levin
  2007). Gravitational collapse of the disk may be especially
  important in star formation bursts close to the
  Galactic centre. Meanwhile, the AGN gas can be relatively well ionised and
  able to couple to any latent magnetic field; indeed the MRI and GI
  may overlap at certain radii in especially luminous systems 
  (Menou \& Quataert 2001).  

These exploratory 2D results point to a number of future research directions. 
For a start, Ohmic diffusion could be replaced by ambipolar diffusion
 to test how magnetic fields
behave in the regimes more relevant for the outer radii of
protostellar disks. However, the most interesting avenues involve 3D
vertically stratified boxes, which could include the $z$-dependent
diffusivities and the various interesting non-ideal
MHD behaviours recently discovered \citep{lesur14,bai14, simon15}. The latter would then provide magnetic fields
self-consistently. Such simulations would let us probe how the
gravitoturbulence works in the presence of MHD winds, surface
turbulence, and its action on the magnetic field. And though
numerically intensive, they would also
provide a way to simulate both the MRI and GI together and determine
if the two instabilities coexist or attempt to switch each other off.

\section*{Acknowledgements}

The authors would like to thank the anonymous reviewer for a helpful
set of comments. They are also indebted to Sijme-Jan Paardekooper and
Charles Gammie for generously reading through an earlier draft and
offering advice and criticism.
This research is partially funded by STFC
grant ST/L000636/1.
Most of the simulations were run on the
DiRAC Complexity system, operated by the University of Leicester
IT Services, which forms part of the STFC DiRAC HPC Facility
(www.dirac.ac.uk). This equipment is funded by BIS National E-
Infrastructure capital grant ST/K000373/1 and STFC DiRAC 
Operations grant ST/K0003259/1. DiRAC is part of the UK National
E-Infrastructure.




\bibliographystyle{mnras}
\bibliography{refs} 


\appendix
\section{Test of the linearised problem}
\subsection{Axisymmetric case (isothermal)}

\begin{figure}[H]
\begin{center}

\label{fig_text_axi}
\includegraphics[width=\columnwidth]{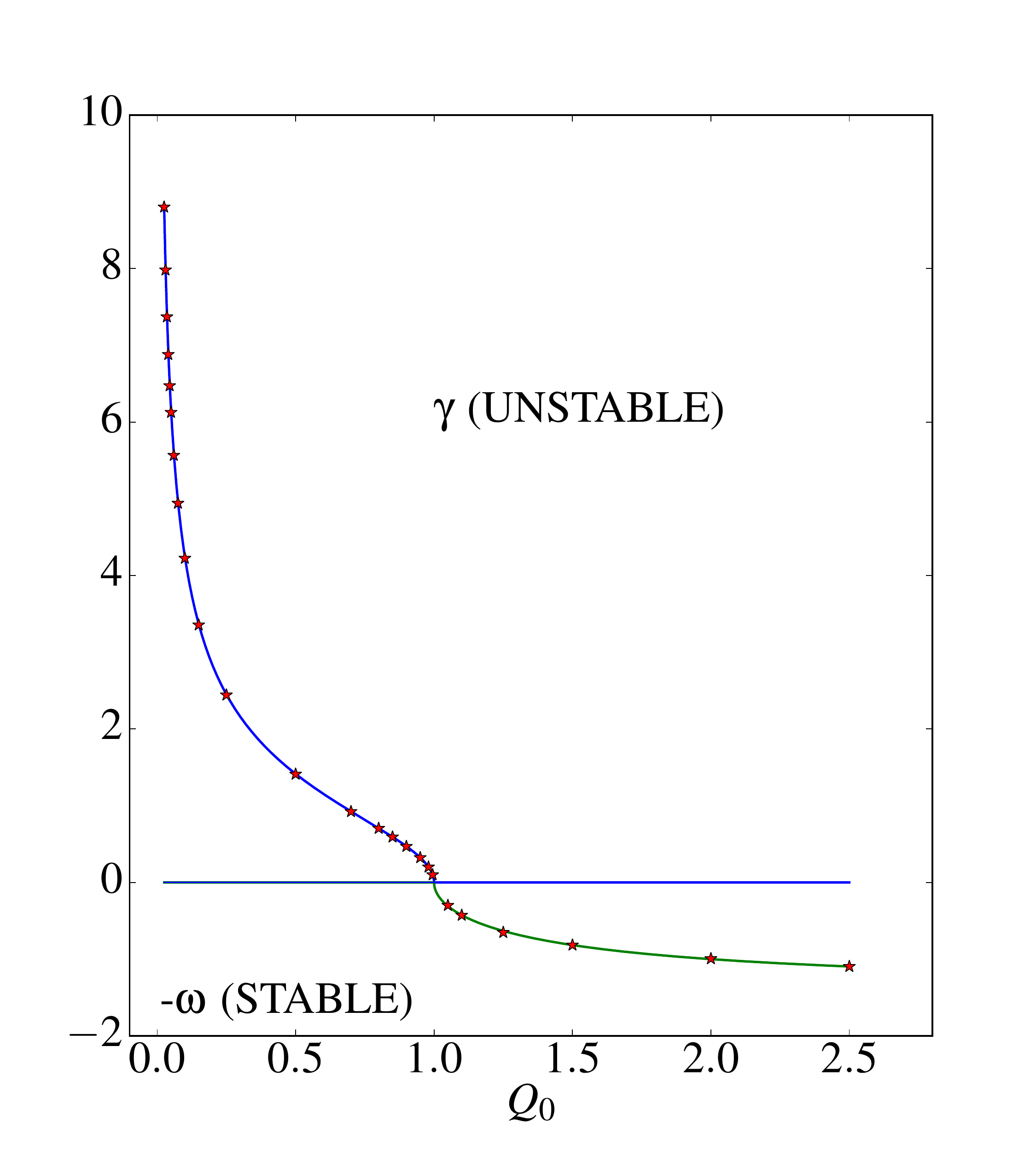} 
\caption{Stability of axisymmetric modes in the local Cartesian thin disc. Solid lines are the theoretical growth rates (positive ordinates) and frequencies (negative ordinates). Red stars are the values obtained with our self-gravity module (working with PLUTO). The resolution is 512 points per azimuthal wavelength.}
\end{center}
\end{figure}

We present in this appendix several tests to check that our self-gravity module in PLUTO is correctly implemented. A first test concerns the linear axisymmetric modes. By linearising the system of equations  \eqref{mass_eq}-\eqref{energy_eq} and assuming that perturbations are of the form $\exp(\text{i}k_xx+\text{i}k_yy-\text{i}\omega t)$, one can derive a dispersion relation which writes, in the isothermal, inviscid and unmagnetized case
\begin{equation}
\label{eq_gamma}
\gamma^2\equiv-\omega^2=-(k_x^2c_s^2+\kappa^2-2k_x\kappa c_s/Q).
\end{equation}
If the Toomre parameter $Q={\kappa c_s}/{\pi G\Sigma_0}<1$, then there exists a range of wavenumbers  for which the motion is unstable.   
\begin{equation}
\dfrac{\kappa}{c_s}\left(\dfrac{1}{Q}-\sqrt{\dfrac{1}{Q^2}-1}\right)\leq k_x\leq\dfrac{\kappa}{c_s}\left(\dfrac{1}{Q}+\sqrt{\dfrac{1}{Q^2}-1}\right)
\end{equation}
To check this relation numerically, we considered a wave in a sheet of
size $L_x= L_y=2\pi$ with a background density $\Sigma_0=1$. We assume
an isothermal gas ($c_{s_0}=1$) and introduce at $t=0$ a perturbation
with wavenumbers $k_x=2 \pi/L_x$ and $k_y=0$ so that it is marginally unstable for $Q=1$. 
We perturbed the background along the dominant eigenvector of the linearised problem, ensuring that the evolution of the wave is strictly exponential when the motion is unstable. The initial amplitude is $10^{-5} S\,H$. We have simulated the evolution of these perturbations for different values of $Q$. Figure \ref{fig_text_axi} shows the numerical growth rates $\gamma$  as a function of $Q$, obtained for $Q<1$ and the frequencies $\omega$ obtained for $Q>1$ (red stars). The blue and green lines represent these quantities obtained analytically from equation \ref{eq_gamma}. The relative error between the theoretical and numerical growth rates remains smaller than 0.008 and on average equal to 0.005.

\subsection{Non-axisymmetric case}

We performed similar tests for non-axisymmetric waves ($k_y\neq 0$). Because these waves have a wavenumber $k_{x}=k_{x_0}+Sk_yt$ that increases linearly in time, analytical solutions are not straightforward to obtain. They are rapidly sheared out and their evolution on long time scale cannot be described by exponentials. However, it is still possible to solve numerically the linearised problem with a simple Runge Kutta time-stepping algorithm and compare the results with the solutions obtained numerically with PLUTO. \\

We considered a leading wave with $k_x=-4\pi/L_x$ and $k_y=2\pi/L_y$ in a box of size $L_x= L_y=2\pi$. We tested 3 different configurations by fixing $Q=1.1358$, $\Sigma_0=1$, $c_{s_0}=1$ and the random initial amplitudes of the velocity and density perturbations.  In the first case, the equation of state is isothermal and the gas is unmagnetized. In the second case, $\mathbf{B}=0$ but the energy equation is taken into account. The last configuration accounts for an ideal and magnetized gas, in which a constant magnetic background $B_x=0.15$ and $B_y=0.3$ is introduced. \\

\begin{figure}[H]
\label{fig_test_naxi}
\begin{center}
\includegraphics[width=\columnwidth]{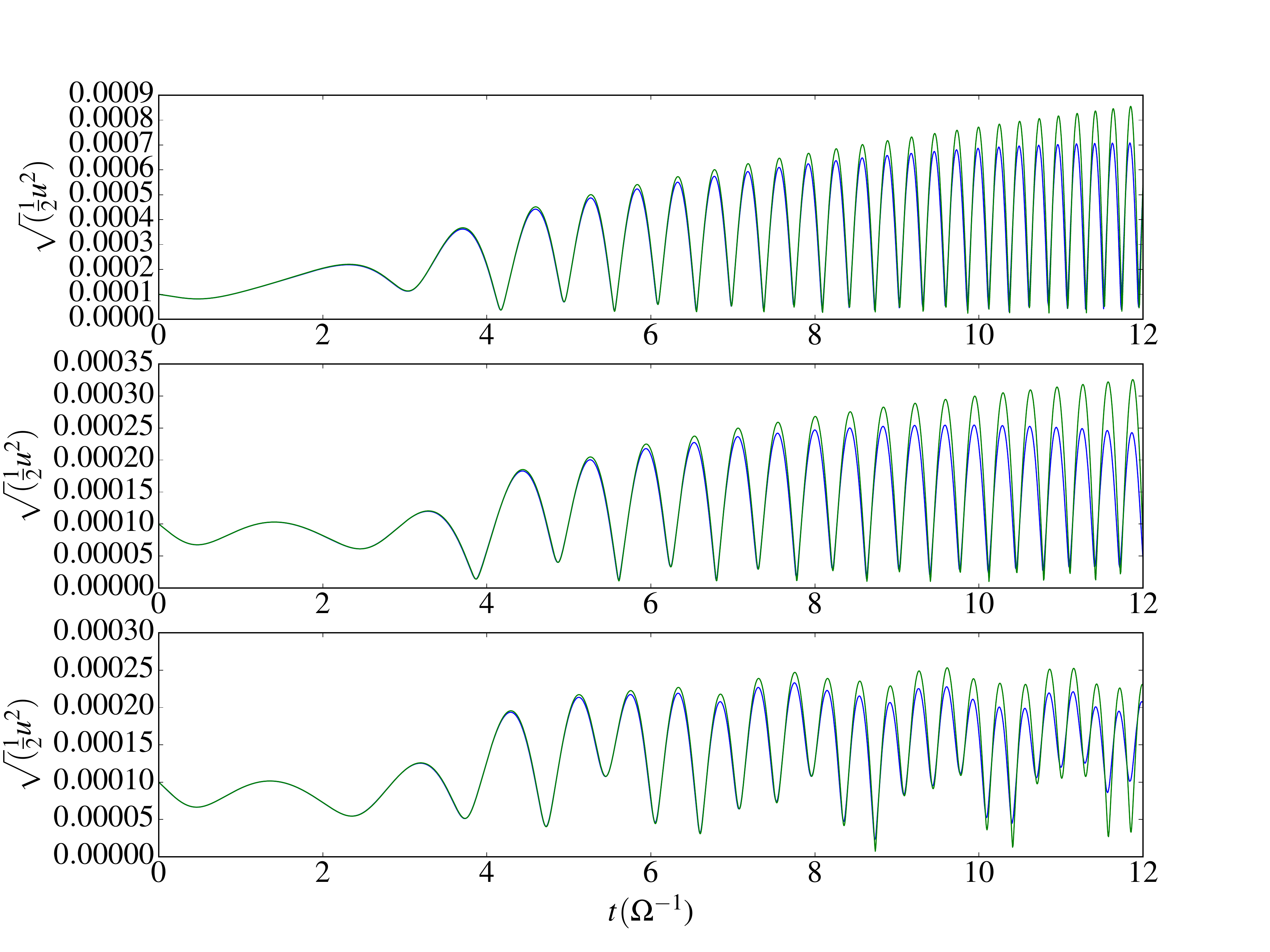} 
\caption{Rms velocity fluctuations of a shearing wave for a
  non-magnetized isothermal gas (top panel), ideal gas (center panel)
  and ideal+magnetized gas (bottom panel). The parameters are $Q=1.1368$, $\Sigma_0=1.0$, $\nu=0$, and $\eta=0$. Green curves represent the theoretical results while blue curves are those obtained with our self-gravity module. The resolution is 512 points per azimuthal wavelength.}
\end{center}
\end{figure}

The results are shown in Fig.~\ref{fig_test_naxi}, where blue curves represent the evolution of shearing waves simulated with PLUTO and the green one with the linearized solver. Our code reproduces quite well the desired solution during the first shearing times but the waves diffuse at longer times ($t>10\,\Omega^{-1}$) as they are strongly sheared out and damped by numerical diffusion. The resolution used here is $512 \times 512$ and we checked that doubling the resolution clearly improves the results. Including an explicit viscosity with Re $ = 1000$ reduces the numerical diffusion. The diffusion of shearing waves by a Godunov scheme has been already pointed out  by \citet{Pdkooper2012}. Note that in his case, for the  best resolution used ($128$ points per wavelength), the amplitude of the linear waves has been reduced by $\sim 10-15\%$ after $t=8 \Omega^{-1}$. In our test simulations, waves are damped by $\sim 5-7\%$ after the same time for the three different configurations.  

We performed the same test for an initial  $Q<1$, but complications
arise in this configuration. Indeed, the code produces artificially
very small leading perturbations at $t\simeq 0$ that are amplified
exponentially when they move from leading to trailing (the `aliasing'
problem). As a result, 
the initial wave is mixed with artifical waves which can potentially
grow faster. After 
a given time, the code fails to reproduce the theoretical
solution. Note that 
filtering these leading modes gives the desired result. \\

\label{test_linear}

\bsp	
\label{lastpage}
\end{document}